\documentclass[
prb,twocolumn,notitlepage,amsmath,amssymb,floats,superscriptaddress,nofootinbib,10pt]{revtex4-2}
\usepackage{graphicx,color}
\usepackage{amsmath}
\usepackage{amsfonts, amsbsy}
\usepackage{amssymb}
\usepackage{eqnarray}
\usepackage{bm}
\usepackage{blkarray}
\usepackage{mathtools}
\usepackage{psfrag}
\usepackage{multirow}
\usepackage{textcomp}
\usepackage{units}
\usepackage{lipsum}
\usepackage[title]{appendix}
\usepackage{soul}
\usepackage{titlesec}
\usepackage{times}
\usepackage{enumitem}
\usepackage{soul}
\usepackage[sort&compress]{natbib}
\usepackage[breaklinks,colorlinks = true,linkcolor = blue,urlcolor  = blue,citecolor = blue,anchorcolor = blue]{hyperref}
\usepackage{float}
\usepackage{orcidlink}

\usepackage{esint}
\usepackage{bbm}

\usepackage{xcolor}

\newcommand{\thint}{%
   \mathop{%
    \mathchoice
      {\!\vcenter{\hbox{\rule[0.6ex]{0.6em}{0.1ex}}}\kern-1.0em\int}
      {\!\vcenter{\hbox{\rule[0.6ex]{0.5em}{0.1ex}}}\kern-1.1em\int}
      {\!\vcenter{\hbox{\rule[0.5ex]{0.4em}{0.08ex}}}\kern-1.0em\int}
      {\!\vcenter{\hbox{\rule[0.4ex]{0.3em}{0.06ex}}}\kern-0.85em\int}%
  }%
}

\newcommand{\R}{\mathcal{R}}
\newcommand{\Kf}{\mathcal{K}_\varphi}

\begin{document}
\title{Geometry-Induced Skin Effect in Electron Hydrodynamics}

\author{Jaros\l{}aw Paw\l{}owski \orcidlink{0000-0003-3638-3966}}
\email{jaroslaw.pawlowski@pwr.edu.pl}
\affiliation{Institute of Theoretical Physics, Wroc\l{}aw University of Science and Technology, 50-370 Wroc\l{}aw, Poland}

\author{Piotr Sur\'owka \orcidlink{0000-0003-2204-9422}}
\email{piotr.surowka@pwr.edu.pl}
\affiliation{Institute of Theoretical Physics, Wroc\l{}aw University of Science and Technology, 50-370 Wroc\l{}aw, Poland}

\author{Konstantin Zarembo \orcidlink{0000-0002-9123-9580}}
\email{zarembo@nordita.org}
\affiliation{Nordita, KTH Royal Institute of Technology and Stockholm University,
Hannes Alfv\'ens v\"ag 12, 106 91 Stockholm, Sweden}
\affiliation{Niels Bohr Institute, Copenhagen University, Blegdamsvej 17, 2100 Copenhagen, Denmark}

\date{\today}

\begin{abstract}
In ultra-clean 2d materials electron viscosity is as important as Ohmic dissipation and electron transport exhibits hydrodynamic features. Using a simple framework of Brinkman equations we find that hydrodynamic electron flows give rise to a geometric skin effect: sharp obstacles locally enhance the current suppressing it far from the edges where the flow is unobstructed. This effect arises within hydrodynamic transport with finite momentum relaxation and does not rely on ballistic dynamics. Our results provide a natural hydrodynamic interpretation of edge-enhanced and double-bump current profiles observed in constricted geometries. By comparing with recent scanning NV magnetometry experiments on gated graphene, we demonstrate that such flow patterns are consistent with viscous hydrodynamics shaped by geometry, clarifying the role of geometric effects in the interpretation of electronic flow experiments.
\end{abstract}

\maketitle

\section{Introduction}

Electron hydrodynamics has emerged as an increasingly important framework for understanding charge transport in ultra-clean conductors. In this regime, momentum-conserving electron–electron collisions dominate over impurity and phonon scattering, allowing the electronic system to behave collectively as a viscous charged fluid \cite{Molenkamp1994,deJong1995,Crossno2016,Bandurin2016,KrishnaKumar2017,berdyugin_measuring_2019,sulpizio_visualizing_2019,ku_imaging_2020,keser_geometric_2021,samaddar_evidence_2021,jenkins2022imaging,krebs_imaging_2023,palm_observation_2024,majumdar_universality_2025}. Such behavior leads to transport signatures that deviate qualitatively from those expected in the conventional Ohmic or ballistic limits \cite{Gurzhi1963,Gurzhi1968,Gurzhi1995} (see \cite{narozhny_hydrodynamic_2022,fritz_hydrodynamic_2024,hui_hydrodynamics_2025} for recent reviews).

Among the various geometries used to probe these effects, constrictions offer a particularly sensitive setting \cite{Levitov2016,KrishnaKumar2017}. Unlike extended channels, constrictions reveal flow patterns that strongly depend on the interplay between geometry and microscopic scattering lengths. Recent scanning NV magnetometry experiments on high-mobility graphene have taken advantage of this sensitivity, directly imaging the current distribution as the system crosses over from Ohmic to non-Ohmic transport regimes \cite{ku_imaging_2020,jenkins2022imaging}. A pronounced double-bump profile observed at room temperature gradually evolves into a single-peak structure at lower temperatures, and is commonly interpreted as reflecting a shift in the balance between momentum-relaxing and momentum-conserving scattering. This interpretation implicitly treats the profile shape as a direct diagnostic of the underlying transport regime. {\color{black}Building upon observation in  \cite{2020PhRvB.102l5404P} we will show that this identification is not unique: similar edge-enhanced structures can arise within a purely hydrodynamic description once the constriction geometry and finite momentum relaxation are taken into account. We refer to this stationary redistribution as a \emph{geometry-induced skin effect}. By this term, we mean the geometrically induced enhancement of the current near sharp boundary features, such as the endpoints of a constriction, accompanied by its relative suppression in the unobstructed interior.}

Previous theoretical analyses of these experiments have relied primarily on kinetic theory simulations, which provide a detailed and flexible description of the crossover between ballistic, hydrodynamic, and Ohmic regimes. While powerful, such approaches tend to entangle geometric effects with microscopic scattering mechanisms, thereby obscuring the hydrodynamic origin of the observed flow patterns and motivating a more transparent continuum description.

In this work, we develop a simpler, continuum hydrodynamic description based on the Brinkman fluid model \cite{brinkman_calculation_1949}, which augments the Stokes equation with a finite momentum-relaxation length. This formulation,  similar in spirit to the holographic model of \cite{Huang:2021mee}, retains the essential ingredients of viscous and Ohmic transport while allowing for transparent analytical treatment of the velocity and potential fields. Following the approach of Falkovich and Levitov \cite{Levitov2016}, the governing equations are solved by Fourier transformation, which reduces the problem to a fourth-order ODE in the transverse velocity. We express the resulting boundary value problem as a singular integral equation for the velocity profile at the constriction, whose structure encodes the transition from narrow-aperture flow to wide-opening behavior and naturally gives rise to the double-bump structure of the current  \cite{2020PhRvB.102l5404P}.

To complement this analytical treatment, we also perform direct numerical simulations of the Brinkman–Navier–Stokes equations in constriction and pipe geometries. The flow is modeled in the low-Reynolds-number regime relevant to electronic hydrodynamics, with no-slip boundary conditions at the walls and controlled variation of the momentum-relaxation rate. By systematically varying geometric scales in the problem, we explore the transition from viscous to Ohmic flow and map how the double-bump structure evolves in shape and amplitude.

The numerical solutions confirm the analytical predictions in the low-Reynolds limit and extend them into intermediate regimes that are analytically inaccessible. Together, the analytical Brinkman model and numerical simulations provide a coherent, minimal description of flow through constrictions, capturing both the geometric origin of the observed flow patterns and their evolution under momentum relaxation. This approach offers a transparent physical interpretation and serves as a natural baseline for more elaborate kinetic or microscopic treatments of electronic transport.

\section{Brinkman equations}

The hydrodynamic description of electron transport is a simple extention of Ohm's law by a diffusion term representing viscosity of the electronic fluid:
\begin{equation}
-\lambda ^2\nabla^2 j_i + j_i = \sigma \partial_i \varphi,
\qquad \partial_i j^i = 0.
\label{eq:brinkman}
\end{equation}
Here and below, Latin indices $i,j\in\{x,y\}$ label Cartesian spatial components, repeated indices are summed, and $\partial_i\equiv\partial/\partial x^i$. Since the spatial metric is Euclidean, indices are raised and lowered with $\delta_{ij}$. The quantity $j_i$ is the two-dimensional electric-current density, $\phi$ is the electrostatic potential, $\sigma$ is the sheet conductivity, and $\lambda$ is the momentum-relaxation length that controls the relative importance of viscous momentum diffusion and Ohmic relaxation. The second equation in Eq.~(1) expresses local charge conservation in the stationary state.

In fluid mechanics Eqs.~(\ref{eq:brinkman}) are known as Brinkman (or Darcy-Brinkman) equations.  A possible, albeit dubious argument for their relevance to electronic transport is that the first equation simply aloows for the {\it spacial dispersion} of conductivity \cite{tanner_optical_2019}: $\sigma(k)=\sigma/(1+\lambda^2k^2)$, while the second equation reflects charge conservation. 

A more refined reasoning relies on the constitutive relations for  the stress-tensor of the electronic fluid and results in the Navier-Stokes equations. The Brinkmann equations arise in the static overdamped approximation, see~\cite{Lucas2018} for a review of this approach. When the current is expressed through velocity and the potential through pressure: $j_i=-env_i$, $P=en\varphi $ ($n$ is the charge carrier density), the Brinkmann equations assume the standard hydrodynamic form:
\begin{equation}
 \eta \,\partial ^2v_i-\frac{\eta }{\lambda ^2}\,v_i=\partial _iP,
\end{equation}
where the viscosity $\eta$, more familiar in the hydrodynamics, is traded for the momentum relaxation length  and conductivity in the Brinkmann framework:
\begin{equation}\label{viscosity}
 \eta =\frac{e^2n^2\lambda ^2}{\sigma }\,.
\end{equation}
Here, $e>0$ denotes the magnitude of the elementary charge, $n$ is the two-dimensional carrier density, $v_i$ is the hydrodynamic velocity, and $P$ is the electronic pressure. With these conventions, the electric current carried by electrons is $j_i=-en v_i$. 


Elementary kinetic-theory estimates give $\eta\sim n m_{\mathrm{e}}v_{\mathrm{e}}\ell_{\mathrm{ee}}$ and $\sigma\sim e^2n\ell_{\mathrm{imp}}/(m_{\mathrm{e}}v_{\mathrm{e}})$. Here, $m_{\mathrm{e}}$ denotes the effective inertial mass of the charge carriers, $v_{\mathrm{e}}$ is their characteristic velocity, and $\ell_{\mathrm{ee}}$ and $\ell_{\mathrm{imp}}$ are the mean free paths associated with electron--electron and momentum-relaxing scattering, respectively. For doped graphene, one may identify $v_{\mathrm{e}}=v_{\mathrm{F}}$ and $m_{\mathrm{e}}=p_{\mathrm{F}}/v_{\mathrm{F}} =\mu/v_{\mathrm{F}}^2$, where $p_{\mathrm{F}}$, $v_{\mathrm{F}}$, and $\mu$ are the Fermi momentum, Fermi velocity, and chemical potential, respectively. These estimates yield $\lambda^2\sim\ell_{\mathrm{ee}}\ell_{\mathrm{imp}}$.

We prefer to view the Brinkmann formulation as a simple phenomenological model, a minimal extension  of the Ohm's law that naturally interpolates between conventional Ohmic flow  at $\lambda=0$ dominated by momentum relaxation and 
purely viscous Stokes flow at $\lambda \to \infty$. In the context of electronic hydrodynamics, this corresponds to a crossover between the dissipative Ohmic regime, where momentum relaxes sufficiently fast on the atomic scale and the Gurzhi regime, where electron--electron collisions dominate and momentum is effectively conserved.

\begin{figure}[t]
 \centerline{\includegraphics[width=7.5cm]{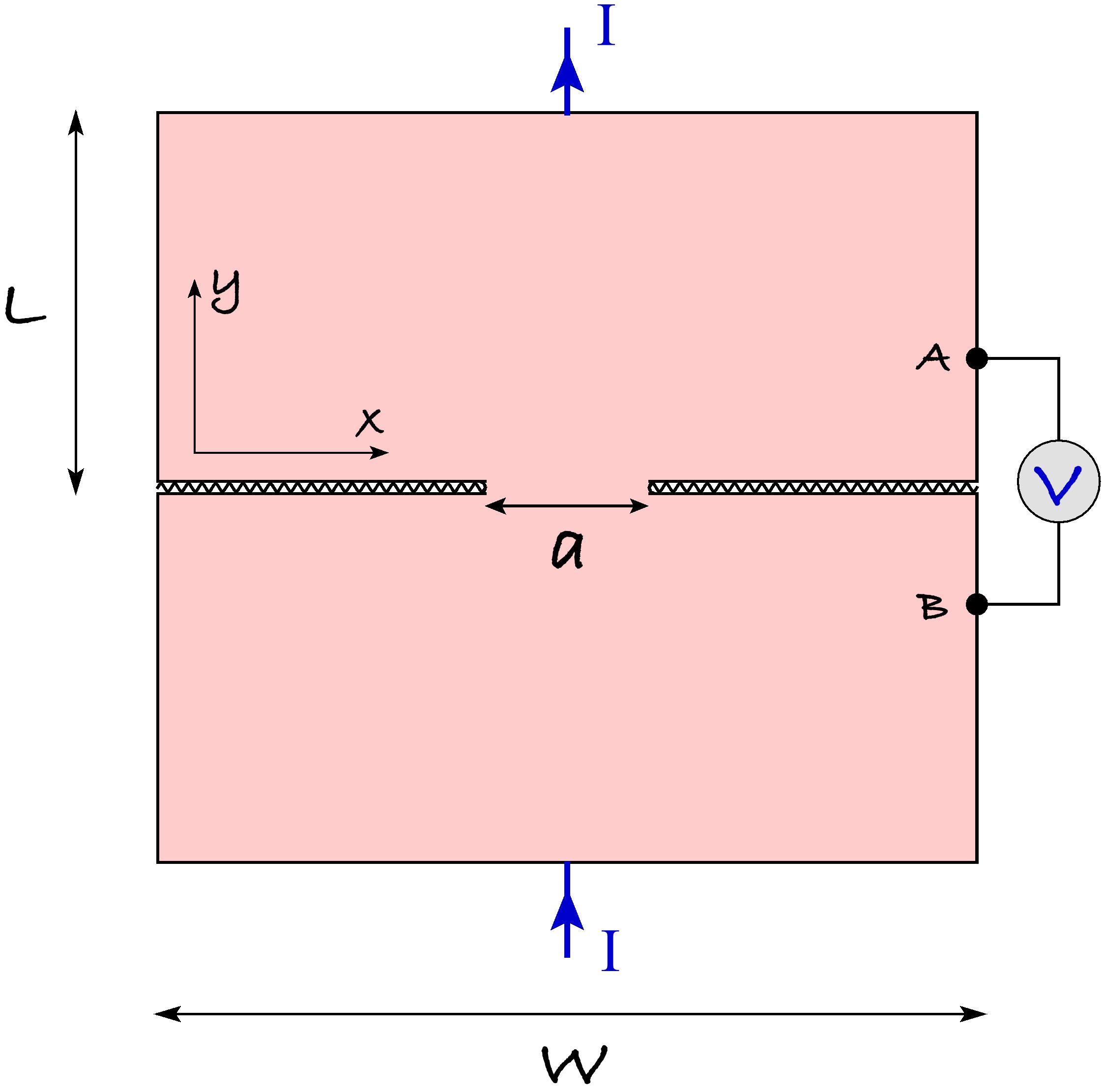}}
\caption{\label{FlowGeo}\small The flow geometry.}
\end{figure}

\section{Brinkman flow through a constriction: Analytic Results}

The hydrodynamic transport is mostly found in 2d materials, and 
we focus on a planar, steady-state flow through a constriction which models typical experimental geometries used in scanning magnetometry and transport measurements.   The current is injected through a point contact into a square reservoir of size $W\times 2L$, split in two halves by a constriction with an opening of width $a$.
The setup is illustrated in Fig.~\ref{FlowGeo}. 

For an infinite sample (that is $W,L\gg a$) the
Brinkmann equations were solved  in \cite{2020PhRvB.102l5404P}. We extend the analysis to finite geometries and in addition calculate the resistance, defined as a potential drop across the constriction divided by the total current: $R=V/I$, see  Fig.~\ref{FlowGeo}. Later we will numerically analyze the influence of different boundary conditions and  the convection term in the hydrodynamic equations.

\subsection{Flow pattern}

In the geometric setup  of fig.~\ref{FlowGeo} the boundary conditions are imposed at $y = 0$: \begin{equation}\label{boundary-out}
 j_i(x,0)=0\qquad  \text{for } \qquad |x| > \frac{a}{2}.
\end{equation}
Impermeability constrains the normal component of the current but does not determine its tangential component. The latter therefore requires an
independent boundary condition. For the analytical calculation, we impose the no-slip condition, which is qualitatively consistent with the strong
suppression of the current near the sample boundaries observed experimentally. In Sec. \ref{maxwellbc}, we relax this assumption and examine
Maxwell partial-slip boundary conditions numerically.

The Dirichlet conditions of this type are commonly called no-slip since they forbid the flow right at the edge of the sample. The no-slip conditions are  qualitatively consistent with the available experimental data. The magnitude of the current, as measured by the NV imaging in \cite{jenkins2022imaging}, indeed drops considerably towards the edges of the constriction. Alternatives, such as no-stress, allow for an unimpeded flow along the edges. The mixed boundary conditions that interpolate between the no-slip and no-stress introduce the new length scale (the slip length) which is argued to become relevant in graphene at low temperatures \cite{KiselevSchmalian,PhysRevB.111.125145}. We will investigate the influence of more general boundary conditions later numerically. To enable the analytic treatment we momentarily set $W=\infty $, returning to effects of a finite width when we turn to numerical solutions. For now on we assume that the $x$ coordinate  varies from $-\infty $ to $+\infty $.

The electron velocity at midflow, directly at the aperture, obviously has no transverse component and
\begin{equation}\label{boundary-in-x}
  j_x(x,0)=0\qquad {\rm for~all~}x.
\end{equation}
The longitudinal  component, on the contrary, is non-zero  and
we introduce a special notation for the midflow current profile:
\begin{equation}\label{boundary-in-y}
j_y(x,0)\equiv I J(x) 
 \quad \text{for } ~~|x| < \frac{a}{2}.
\end{equation}
The shape function
$J(x)$ has to be determined self-consistently from the Brinkmann equations. Conservation of charge fixes its overall normalization:
\begin{equation}\label{normalization-J}
 \int\limits_{-a/2}^{a/2}dx\, J(x)=1.
\end{equation}
The boundary conditions at the opposite side correspond to an outflow through the point contact:
\begin{equation}\label{other-side-conditions}
 j_x(x,L)=0,\qquad j_y(x,L)=I\delta (x).
\end{equation}

There is no need to solve the equations for $y<0$, the flow pattern in the lower half-plane can be reconstructed by symmetry:
\begin{align}\label{symmetryaty=0}
 &j_x(x,-y)=-j_x(x,y)
\nonumber \\
& j_y(x,-y)=j_y(x,y)
\nonumber \\ 
&\varphi (x,-y)=-\varphi (x,y).
\end{align}
With this construction, the potential will acquire a jump across $y=0$. And indeed the potential  need not be the same on the two sides of the constriction: its jump at infinity is precisely the voltage between contacts $A$ and $B$ in fig.~~\ref{FlowGeo}, which  defines the resistance of the whole constriction, and which we will later calculate. The potential at the aperture however must be continuous, allowing for an unimpeded flow between $x=-a/2$ and $x=a/2$. The only way to achieve this without compromising the symmetry (\ref{symmetryaty=0}) is to require the potential to be zero:
\begin{equation}\label{boundary-in-phi}
 \varphi (x,0)=0 \quad \text{for } ~~ |x| < \frac{a}{2}.
\end{equation}
This makes the aperture an equipotential line and, in mathematical terms, supplies the remaining boundary condition to close the system of equations.

In this geometric setup, the Brinkmann equations can be solved by means of partial Fourier transform:
\begin{align}
 j_x(x,y)&=\int\limits_{0 }^{+\infty }\frac{dk}{\pi }\,\, j_x(k,y)\sin kx
 \\
  j_y(x,y)&=\int\limits_{0 }^{+\infty }\frac{dk}{\pi }\,\, j_y(k,y)\cos kx
 \\
 \varphi (x,y)&=\int\limits_{0 }^{+\infty }\frac{dk}{\pi }\,\, \varphi (k,y)\cos kx.
\end{align}
Here we take advantage of an infinite lateral extent of the sample, which is of course an idealization. The numerical results for truly finite geometries presented later indicate that the influence of the remote edges on the solution is negligible, and the infinite-domain treatment accurately captures the local pattern of the charge flow.

Upon the Fourier transform the Brinkmann equations reduce to a set of second-order ODEs:
\begin{align}
& \lambda ^2\left(-\frac{d^2}{d^2y}+\omega ^2\right)j_x=-\sigma k\varphi 
  \\
&  \lambda ^2\left(-\frac{d^2}{d^2y}+\omega ^2\right)j_y=\sigma \,\frac{d\varphi }{dy}\\
  &kj_x+\frac{dj_y}{dy}=0,
\end{align}
where  we introduced a shorthand notation:
\begin{equation}
 \omega =\sqrt{k^2+\frac{1}{\lambda ^2}}\,.
\end{equation} 
Crucially, the two unknown functions $j_x$ and $\varphi $ appear in the first and last equations without derivatives and can be expressed 
algebraically through $j_y$:
\begin{align}
\label{jx-interm}
 & j_x=-\frac{1}{k}\,\,\frac{dj_y}{dy}
 \\
 \label{phi-interm}
 &\varphi =\frac{\lambda ^2}{\sigma k^2}\left(-\frac{d^2}{d^2y}+\omega^2\right)\frac{dj_y}{dy}\,.
\end{align}
The middle equation then yields a fourth-order ODE for the longitudinal component of the current:
\begin{equation}
\left( -\frac{\partial^2}{\partial y^2} + k^2 \right)
\left( -\frac{\partial^2}{\partial y^2} + \omega^2 \right)
j_y = 0.
\label{eq:brinkman-ode}
\end{equation}

The general solution of this equation contains four integration constants, to be fixed later by the boundary conditions:
\begin{equation}\label{jy-solution}
 j_y=I\left(A\cosh ky+B\sinh ky+C\cosh\omega y+D\sinh\omega y\right).
\end{equation}
The transverse component can be inferred from (\ref{jx-interm}):
\begin{align}\label{jx-solution}
 j_x=-\frac{I}{k}&\left(
 kA\sinh ky+kB\cosh ky+\omega C\sinh\omega y
 \right.
\nonumber \\
&\left.
 +\omega D\cosh\omega y
 \right),
\end{align}
and for the potential we get:
\begin{equation}\label{phi-solution}
 \varphi =\frac{I}{\sigma }\int\limits_{0}^{\infty }
 \frac{dk}{\pi k}\,\,
 \left(A\sinh ky+B\cosh ky\right)\cos kx+\varphi _0.
\end{equation}
The integration constants $A,B,C,D$ are all functions of the wave number. On the contrary the integration constant $\varphi_0$ is just a number. The ground level of the potential can be always shifted without affecting the gradients.  

The boundary conditions (\ref{boundary-in-x}), (\ref{boundary-in-y}) and (\ref{other-side-conditions}) take the following form in the Fourier space:
\begin{align}
 &j_x(k,0)=0,\qquad j_y(k,0)=IJ(k)
 \\
 &j_x(k,L)=0,\qquad j_y(k,L)=I.
\end{align}
Substituting the solution obtained above yields
a system of four linear equations for the four integration constants:
\begin{align}
 &kB+\omega D=0
\nonumber \\
&A+C=J
\nonumber \\
&kA\sinh kL+kB\cosh kL+\omega C\sinh\omega L+\omega D\cosh\omega L=0
\nonumber \\
&A\cosh kL+B\sinh kL+C\cosh\omega L+D\sinh\omega L=1.
\end{align}
The solution of this linear system is
\begin{widetext}
\begin{align}
&A=\omega \,\frac{\left[
(\omega -k)\cosh(\omega +k)L-(\omega +k)\cosh(\omega -k)L+2k
\right]J-2k\left(\cosh\omega L-\cosh kL\right)}
{(\omega -k)^2\cosh(\omega +k)L-(\omega +k)^2\cosh(\omega -k)L+4\omega k}
\nonumber \\
\label{thatsbe}
&B=-\omega \,\frac{\left[
(\omega -k)\sinh(\omega +k)L+(\omega +k)\sinh(\omega -k)L
\right]J-2\omega \sinh\omega L+2k\sinh kL}
{(\omega -k)^2\cosh(\omega +k)L-(\omega +k)^2\cosh(\omega -k)L+4\omega k}
\nonumber \\
&C=J-A
\nonumber \\
&D=-\frac{k}{\omega }\,B.
\end{align}
\end{widetext}
These equations, together with (\ref{jy-solution}), (\ref{jx-solution}) and (\ref{phi-solution}), express the velocity field and potential in terms of a single function $J(k)$, the Fourier transform of the velocity profile at the aperture. To find this function we need to impose the last remaining condition, which requires the potential to be constant across the aperture.

\subsection{Integral equation for flow profile}

In terms of the explicit solution (\ref{phi-solution}), the condition
of vanishing potential (\ref{boundary-in-phi}) becomes
\begin{equation}
 \frac{I}{\sigma }\int\limits_{0}^{\infty }\frac{dk}{\pi k}\,\,
 B\cos kx+\varphi _0=0,
\end{equation}
and this should hold on the interval $-a/2<x<a/2$. Ostensibly arbitrary constant $\varphi _0$ is in fact fixed by the equation itself. The ground level of the potential can of course be chosen at will, but here we insist on the symmetry (\ref{symmetryaty=0}) and it only holds for one particular choice. Dealing with an implicitly defined constant of no physical significance is not very convenient, and to get rid of it we differentiate the last equation in $x$:
\begin{equation}\label{B-equation-Fourier}
 \int\limits_{0}^{\infty }\frac{dk}{\pi}\,\,
 B\sin kx=0.
\end{equation}

The constant $B$ is a fairly complicated expression that depends on $k$ and contains an unknown function $J(k)$. A closer inspection of (\ref{thatsbe}) reveals that $B$ has a term linear in $J$ and a free term fully specified. Splitting the two and transforming back into the coordinate representation we get from  (\ref{B-equation-Fourier})  a linear integral equation for the current profile:
\begin{equation}\label{main-integral-equation}
 \fint\limits_{-a/2}^{a/2}dx'\,K(x-x')J(x')=r(x)\qquad \text{for } ~~ |x| < \frac{a}{2}\,.
\end{equation}
The kernel and the driving term are given by
\begin{widetext}
\begin{align}
&
\label{kernel-def}
K(x) =
\int\limits_{0}^{\infty }
\frac{dk\,\omega }{\pi }\,\,
\frac{
(\omega -k)\sinh(\omega +k)L+(\omega +k)\sinh(\omega -k)L}
{(\omega -k)^2\cosh(\omega +k)L-(\omega +k)^2\cosh(\omega -k)L+4\omega k}
\,\sin kx
\\
&
\label{driving-term-def}
r(x)=2
\int\limits_{0}^{\infty }
\frac{dk\,\omega }{\pi }\,\,
\frac{\omega \sinh\omega L-k\sinh kL}
{(\omega -k)^2\cosh(\omega +k)L-(\omega +k)^2\cosh(\omega -k)L+4\omega k}
\,\sin kx.
\end{align}
\end{widetext}

These functions cannot calculated in the closed form, but can be easily computed numerically. It is also easy to establish their general analytic properties:
\begin{itemize}
 \item $K(x)$ and $r(x)$ are odd functions of $x$.
 \item At $x\rightarrow \infty $, $K(x)$ and $r(x)$ asymptote to the same positive constant:
\begin{equation}\label{large-x-Kr}
 K(x),\, r(x)\stackrel{x\rightarrow \pm\infty }{\simeq }
 \pm\frac{1}{2L-4\lambda \tanh\frac{L}{2\lambda }}\,.
\end{equation}
The next term is $\mathcal{O}(1/x^2)$ which will be important in the computation of resistance.
 \item The function $r(x)$ is completely regular.
 \item The kernel, on the contrary, is singular at zero:
\begin{equation}
 K(x)\stackrel{x\rightarrow 0}{\simeq }-\frac{4\lambda ^2}{\pi x^3}+\frac{3}{2\pi x}+{\rm finite}.
\end{equation}
A convolution with $K$ therefore requires regularization.
In the integral equation (\ref{main-integral-equation}) the divergence is treated with the principal-value prescription:
\begin{equation}
 \mathop{\mathrm{P.V.}}K(x)\equiv \frac{K(x+i\epsilon )+K(x-i\epsilon )}{2}\,.
\end{equation}
One can argue on general grounds that the principal value regularization is the only one consistent with symmetries. A more solid argument involves returning to (\ref{phi-solution}) and evaluating the potential at small but non-zero $y$. A   lengthy  but straightforward calculation shows that the limit $y\rightarrow 0$ is equivalent to the principal-value regularization with $y$ playing the role of $\epsilon $.
\end{itemize}

The flow profile therefore is a solution to a {\it singular} integral equation (\ref{main-integral-equation}). Integral equations of this type are ubiquitous in random matrix theory, Bethe ansatz and other areas of mathematical physics. A well-developed analytic theory \cite{Gakhov,muskhelishvili2013singular} guarantees existence and uniqueness of the solution with zero boundary conditions at the endpoints, where the solution behaves as
\begin{equation}\label{true-behavior-edges}
 J(x)\stackrel{x\rightarrow \pm a/2}{\sim }\sqrt{a^2-4x^2}\,.
\end{equation}
We stress that the square-root scaling is not imposed by hand but is mandated by the equation itself. Uniqueness of the solution requires clarification. Because the kernel has a $1/x^3$ singularity, the integral equation itself has a one-parametric family of solutions, the parameter is then fixed by the normalization condition (\ref{normalization-J}). 

Once the flow profile at the aperture is determined from 
(\ref{main-integral-equation}), the full solution in the $(x,y)$ plane can be reconstructed from (\ref{jy-solution}), (\ref{jx-solution}), (\ref{phi-solution}) and (\ref{thatsbe}). While in general this can be only done numerically, further analytical progress is possible in several limiting cases which we explore below.

\section{Resistance}

The resistance of the constriction is defined through the potential difference between the points $A$ and $B$ in fig.~\ref{FlowGeo}:
\begin{equation}
    R=\frac{\varphi_A-\varphi_B}{I}\,.
\end{equation}
In the figure the points are split for visual clarity. In practice we assumed that the contacts are placed right after and right before the constriction at the extremity of the sample. 

\subsection{Order-of-magnitude estimates}
\label{estimates-of-resistance}

According to basic textbook definitions the resistance is proportional to the length and inverse proportional to the cross-section of the sample and hence the most naive estimate for the geometry at hand is
\begin{equation}\label{classR}
    R\sim \frac{L}{\sigma a}\,.
\end{equation}
This simplistic guess appears to be a gross overestimate. 

At the other extreme  the fully quantum regime of ballistic transport in, when the resistance is set by the quantum unit of resistivity  $\hbar/e^2$ and  the size of the Fermi surface \cite{Sharvin1965,vanHoutenBeenakker1996}:
\begin{equation}
 R_{\rm ball}\sim \frac{\hbar^2}{e^2am_ev_e}\,.
\end{equation}
Translating this expression in the kinetic-theory language by taking $\sigma \sim e^2n\ell_{\rm imp}/m_ev_e$ for the resistivity and $n\sim p_F^2/\hbar^2\sim m_e^2v_e^2/\hbar^2$ for the carrier density, we get:
\begin{equation}
 R_{\rm ball}\sim \frac{l_{\rm imp}}{\sigma a}\,.
\end{equation}

This is certainly smaller than the classical resistance (\ref{classR}).
But, as pointed out in  \cite{Guo2017higher},
the hydrodynamic resistance can be even smaller, and this superballistic 
behavior was indeed observed experimentally \cite{KrishnaKumar2017}. Indeed, in the regime of low momentum dissipation $\lambda \gg a$, the hydrodynamic resistance scales as
\begin{equation}
 R_{\rm hydro}\sim \frac{\lambda ^2}{\sigma a^2}\,.
\end{equation}
The hydro transport is therefore superballistic when $\sqrt{l_{\rm imp}l_{ee}}\gg a\gtrsim l_{ee}$ (where we used that $\lambda ^2\sim l_{\rm imp}l_{ee}$), which requires a clean sample and a relatively large aperture in the constriction.

\subsection{Hydrodynamic resistance}

We can calculate the hydrodynamic resistance very precisely in the Brinkman framework. It is convenient to represent the voltage as an integral of the potential gradient:
\begin{equation}
    \varphi_A-\varphi_B=\int\limits_B^Adx^i\,\partial_i\varphi.
\end{equation}
The natural contour of integration wraps one half of the constriction passing through the aperture. The potential gradient on this contour for the solution (\ref{phi-solution}) follows from the symmetry relations (\ref{symmetryaty=0}) which implies that
\begin{equation}
 \left.\frac{\partial \varphi }{\partial x}\right|_{y=\pm \varepsilon }
 =\mp\frac{I}{\sigma }\int\limits_{0}^{\infty }\frac{dk}{\pi }\,\,B\sin kx.
\end{equation}
It is convenient to introduce a function:
\begin{equation}
 \R(x)=-\int\limits_{0}^{\infty }\frac{dk}{\pi }\,\,B\sin kx,
\end{equation}
naturally interpreted as the differential resistance.
The full resistance is computes as its integral:
\begin{equation}\label{int-representation-resistance}
 R=\frac{2}{\sigma }\int\limits_{x_0}^{\infty }dx\,\R(x).
\end{equation}

The coordinate-space representation for the differential resistance follows from
the explicit form of the coefficient $B$ in (\ref{thatsbe}) and the definitions (\ref{kernel-def}), (\ref{driving-term-def}):
\begin{equation}\label{differential-resistance}
 \R(x)=\fint\limits_{-a/2}^{a/2}dx'\,K(x-x')J(x')-r(x).
\end{equation}
The integral equation (\ref{main-integral-equation})  implies that $\R(x)=0$ for any $x\in(-a/2,a/2)$. As a consequence, the integral in (\ref{int-representation-resistance})  does not depend on the initial point $x_0$ provided that it lies within the interval $(-a/2,a/2)$. Two convenient choices are $x_0=a/2$ or $x_0=0$.  The differential resistance decreases as $\R(x)\sim 1/x^2$ at large $x$, because of (\ref{large-x-Kr}) and the normalization condition (\ref{normalization-J}), and this property guarantees the convergence of the integral defining the total resistance.

The developed formalism defines the flow pattern and the resistance of the constriction in terms of the solution to an integral equation (\ref{main-integral-equation}), whose kernel  (\ref{kernel-def}) and the driving term (\ref{driving-term-def}) are quite complicated functions for generic $a$, $L$ and $\lambda $. We consider several limiting cases where the solution can be found analytically or semi-analytically.

\section{Large sample }
\label{sec:large-L}

First we consider the best physically motivated regime, where the 
sample is much larger than the aperture of the constriction and the momentum dissipation length: : $L\gg a,\lambda$, making no assumptions on the relative size of $a$ and $\lambda$.
The integral equation (\ref{main-integral-equation}) considerably simplifies when the sample is very large, larger than any other scale in the problem. The source term (\ref{driving-term-def}) then goes to zero as $1/L$ and can be dropped, while the kernel (\ref{kernel-def}) becomes
\begin{equation}\label{large-L-kernel}
 K(x)=\lambda ^2\int\limits_{0}^{\infty }\frac{dk}{\pi }\,\omega (\omega +k)\sin kx.
\end{equation}
This integral can be  expressed through the modified Bessel function of the second kind:
\begin{equation}\label{explicit-kernel}
 K(x)=\frac{1}{\pi x}\mathcal{K}\left(\frac{|x|}{\lambda }\right),
 \qquad \mathcal{K}(s)=-\frac{2}{s^2}+1-K_2(s).
\end{equation}
The outflow into an infinite sample is thus described by a homogeneous equation
\begin{equation}\label{integral-eq-largeL}
 \fint\limits_{-a/2}^{a/2}dx'\,K(x-x')J(x')=0
\end{equation}
with the above kernel.

The flow pattern also considerably simplifies when the sample is very large  
\cite{Zarembo:2019znp}:
\begin{align}
&j_x=I\lambda ^2\,
\omega (\omega +k)\left(\,{\rm e}\,^{-k y}-\,{\rm e}\,^{-\omega y}\right)J(k)
\label{Fouriervx}
\\
&j_y=I\lambda ^2\,
(\omega+k)
\left(\omega \,{\rm e}\,^{-ky}-k\,{\rm e}\,^{-\omega y}\right)J(k)
\label{Fouriervy}
\\
\label{phi-Fourier}
&\varphi =-\frac{I\lambda ^2\,\omega (\omega +k)}{\sigma k}\,{\rm e}\,^{-ky} J(k). 
\end{align}
These formulas can be derives by expanding the general solution 
at large-$L$, but they can 
can be also obtained by directly solving the differential equation (\ref{eq:brinkman-ode}) and demanding that the current dissipates at infinity, which amounts to keeping only the decaying exponentials.
 
The potential acquires a remarkable mathematical feature in this approximation: the $x$ and $y$ coordinates combine under the Fourier integral into a single complex variable
\begin{equation}
 z=x+iy,
\end{equation}
and the electrostatic potential becomes a harmonic function. The inverse Fourier transform of (\ref{phi-Fourier}) indeed gives:
\begin{equation}\label{convolution-P}
 \varphi= \varphi_0  -\frac{\lambda ^2I}{2\pi \sigma }
 \int\limits_{-{a}/{2}}^{{a}/{2}}dx'\,\Kf\left(z-x'\right)J(x')+{\rm c.c.},
\end{equation}
where
\begin{equation}\label{Greens}
 \Kf(z)=\int\limits_{0}^{\infty }dk\,\,{\rm e}\,^{ikz}\,\frac{\omega (\omega +k)}{k}\,.
\end{equation}
 
The integral nominally diverges at small $k$. The physical origin of the divergence is easy to understand: an infinite voltage is needed  to propel electrons through an infinite sample, as long as electrons loose momentum through impurity or phonon scattering. The divergence only arises upon taking the $L\rightarrow \infty $ limit, the finite-$L$ expression (\ref{phi-solution}) is well-behaved and convergent. And the potential gradients remain finite even at infinite $L$. The $x$-gradient, for instance, is given by a convergent integral
\begin{equation}
 \frac{\partial }{\partial x}\,\Kf(z)=i\int\limits_{0}^{\infty }dk\,\,{\rm e}\,^{ikz}\,\omega (\omega +k)\,,
\end{equation}
Imposing the condition that $\partial _x\varphi =0$ at the aperture leads to the integral equation (\ref{integral-eq-largeL}) with the kernel (\ref{large-L-kernel}). 

The homogeneous integral equation 
with the kernel (\ref{explicit-kernel}) was studied in detail in 
\cite{2020PhRvB.102l5404P}. While a fully analytic solution cannot be obtained, a useful approximation consists in replacing the Bessel function in (\ref{explicit-kernel}) with its asymptotics at zero: $K_2(s)\approx 2/s^2-1/2$ \cite{2020PhRvB.102l5404P}, which allows for an explicit analytic solution. We do not make this approximation here and will rather rely on systematic expansions in the limiting cases of large and small $\lambda$ complemented by a numerical solution in the intermediate range of $\lambda$.

\subsection{Weak dissipation}

The analytic solution can be systematically constructed when the momentum dissipation length is large compared to opening the constriction: $a\ll \lambda $. The kernel  (\ref{explicit-kernel}) can then be expanded in $x/\lambda $:
\begin{equation}\label{Kexpanded}
 K(x)\stackrel{\lambda \rightarrow \infty }{\simeq } -\frac{4\lambda ^2}{\pi x^3}+\frac{3}{2\pi x}+\ldots 
\end{equation}
Keeping only the leading term, the integral equation (\ref{integral-eq-largeL}) prescribes the flow profile to be a zero mode of the $1/(x-y)^3$ kernel. The zero mode indeed exists and is unique up to normalization:
\begin{equation}\label{LOlambda=inf}
 J(x)\stackrel{\lambda \rightarrow \infty }{\simeq }\frac{4}{\pi a^2}\,\sqrt{a^2-4x^2}\,.
\end{equation}
The midflow therefore has a semicircular shape in the limit of no dissipation  \cite{Guo2017higher}. The integral equation, being homogeneous, does not determine the overall normalization constant, the latter instead is fixed by the normalization condition (\ref{normalization-J}).

The first dissipative correction starts at the second order in $a/\lambda $ and can be easily calculated. The second term in (\ref{Kexpanded}) gives rise to a $x^2/\lambda ^2$ correction to the prefactor of the semicircle, with a fixed relative coefficient. This in turn produces a $\mathcal{O}(a^2/\lambda^2)$ correction to the overall normalization, leading to
\begin{equation}\label{NLOlambda=inf}
 J(x)\stackrel{\lambda \rightarrow \infty }{\simeq }\frac{4}{\pi a^2}
 \left(1-\frac{a^2}{128\lambda ^2}+\frac{x^2}{8\lambda ^2}\right)
 \sqrt{a^2-4x^2}\,.
\end{equation}
The $x^2$ term is positive which means that the current is redistributed towards the edges of the apertures. 

\subsection{Skin effect}

\begin{figure}[bt]
    \centering
    \includegraphics[width=0.86\linewidth]{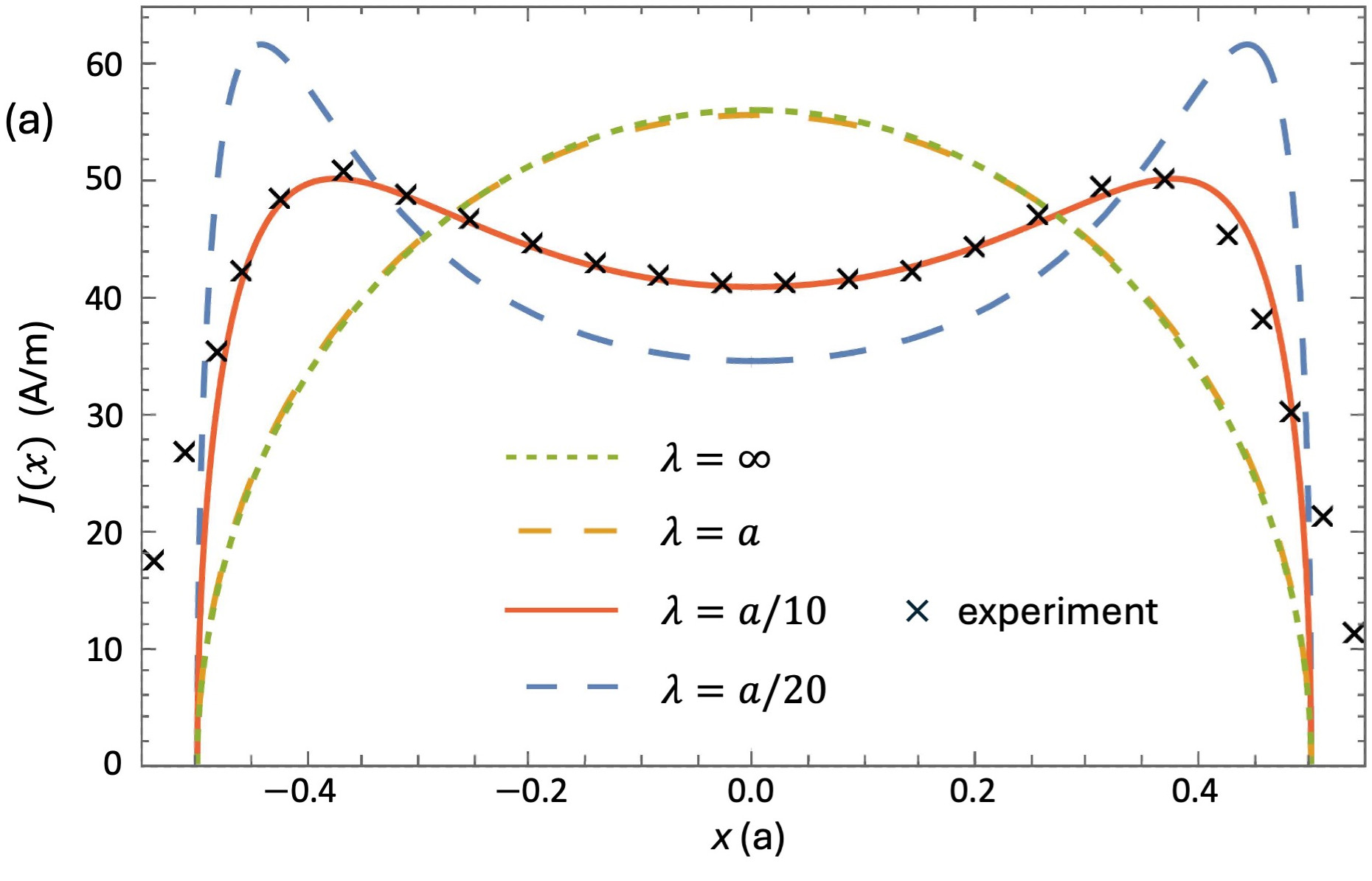}\\
    \includegraphics[width=1.\linewidth]{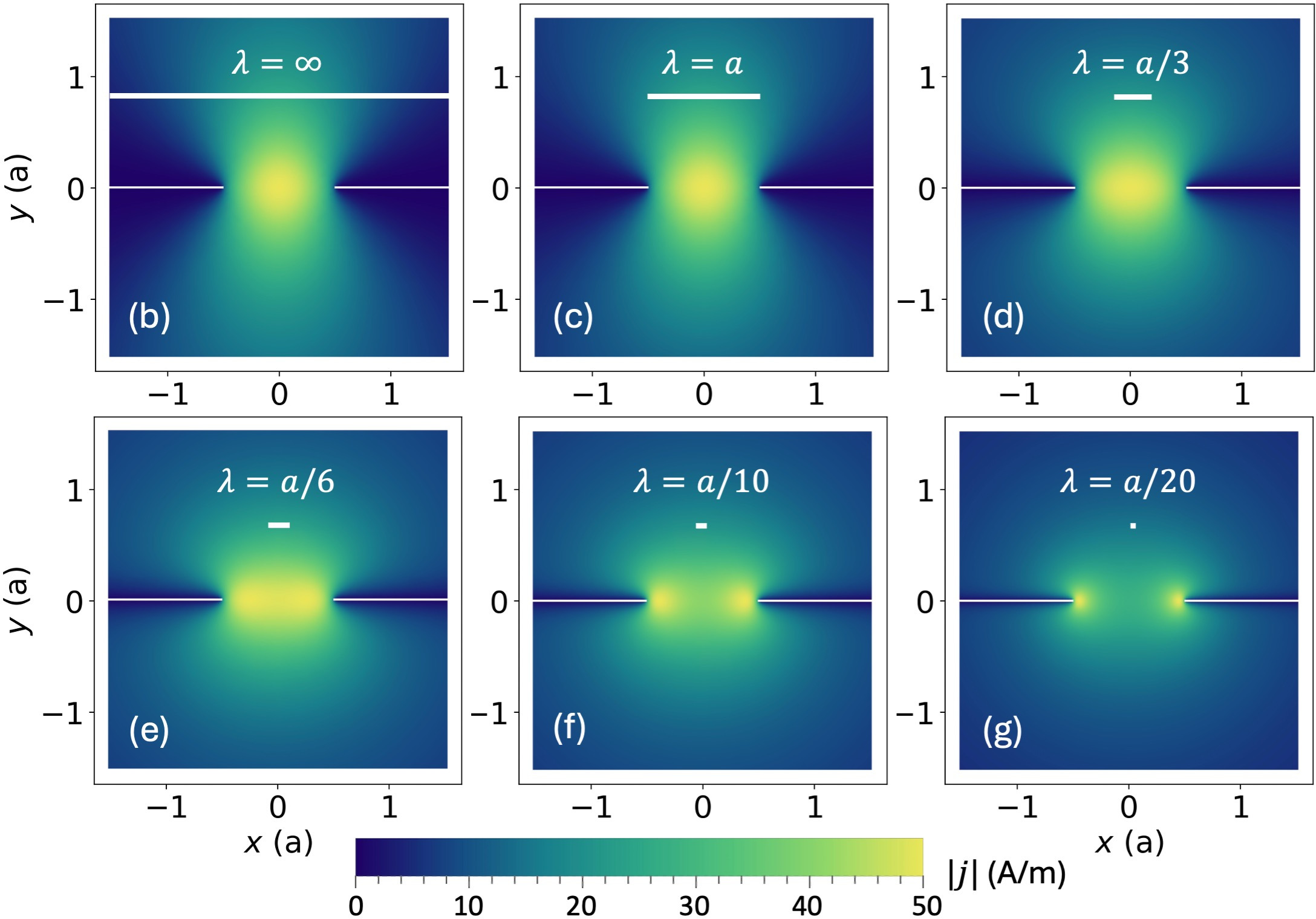}
    \caption{(a) The flow profile at the aperture compared to the experimental data from \cite{jenkins2022imaging}. The solid red line is the best fit with $\lambda=a/10$. The dashed lines correspond to $\lambda=a/20$ (blue) as well as $\lambda=a$ (orange) and $\lambda=\infty$ (green) virtually indistinguishable from one another. (b-g) The current density $|j|$ maps for momentum relaxation length $\lambda=\infty$, $a$, $a/3$, $a/6$, $a/10$, and $a/20$.}
    \label{fig:an}
\end{figure}

This trend continues with growing dissipation and eventually the midflow develops a characteristic double-hump profile  \cite{2020PhRvB.102l5404P} seen in the experiment~\cite{jenkins2022imaging}. This is illustrated in Fig.~\ref{fig:an} which shows the solution of the integral equation, as well as the current density layout reconstructed from (\ref{Fouriervx}) and (\ref{Fouriervy}),  for different values of the dissipation length. 

Taking the divergence and the curl of the Brinkman equation gives, respectively,
\begin{equation}
\nabla^2 \phi = 0,
\qquad
\left(1-\lambda^2\nabla^2\right)\omega = 0,
\qquad
\omega \equiv \epsilon^{ij}\partial_i j_j .
\end{equation}
The current can therefore be decomposed into an irrotational potential sector and a vortical sector screened over the momentum-relaxation length~$\lambda$. In an unconstrained region, the screened contribution decays away from the boundary. Near the sharp endpoints of the constriction, however, impermeability together with the tangential boundary condition requires a rapid rearrangement of the current and excites the screened vortical sector. Matching this sector to the potential flow produces an edge-localized contribution. Consequently, the current is enhanced near the endpoints of the aperture and relatively suppressed in its central region, generating the double-hump profile. The ratio $\lambda/a$ controls this redistribution: for $\lambda\gg a$, the profile approaches the single-peaked Stokes form, whereas decreasing $\lambda/a$ produces progressively stronger edge localization, with the maxima situated at distances of order~$\lambda$ from the aperture endpoints.

The comparison in Fig.~\ref{fig:an}(a) with the experimental profile by Jenkins~\emph{et al.}~\cite{jenkins2022imaging} is a comparison of normalized profile shapes rather than a multi-parameter fit to the complete nitrogen vacancy imaging of narrow graphene
channels data. In particular, we do not attempt to re-extract the microscopic scattering lengths reported in Ref.~\cite{jenkins2022imaging}. Our result demonstrates instead that an edge-enhanced profile of comparable shape can emerge within a Brinkman hydrodynamic model with finite momentum relaxation and remains robust when Maxwell partial-slip boundary conditions are used, as discussed later.

We will call the boundary-localized current enhancement  the “geometry-induced skin effect”. Usually this term  is reserved for the frequency-dependent suppression of an AC current at the electromagnetic skin depth. The two effects should not be confused and they have a different physical origin.
We use the term ``geometry-induced skin effect'' in a specific 
hydrodynamic, rather than electromagnetic, sense. As noted by 
Pozrikidis~\cite{Pozrikidis1992}, after a temporal Fourier or Laplace transform, the 
unsteady Stokes equation involves the same modified Helmholtz 
operator as the Brinkman equation, with the frequency-dependent 
penetration scale replaced by the momentum-relaxation length 
$\lambda$. In time-dependent oscillatory electronic flows, this operator produces a viscous boundary layer and can split the maximum of the flow profile into two ridges localized 
near the boundaries~\cite{moessner_pulsating_2018}, which is a direct analogue of the AC skin effect. The stationary Brinkman 
problem studied here exhibits an analogous spatial localization, but 
one generated by momentum relaxation together with the constriction 
geometry. The phenomenon is therefore distinct from the conventional 
electromagnetic skin effect: it involves neither electromagnetic 
induction nor frequency-dependent screening. We use the qualified 
term ``geometry-induced skin effect'' to emphasize the shared 
screened-Stokes structure and the resulting boundary localization, 
rather than to identify the two physical mechanisms.

The skin effect is clearly visible for strong enough dissipation, the current then flows primarily near the sharp edges of the constriction. In more general geometries the  current tends to be enhanced near sharp edges. We will see some examples when discussing numerical solutions of the Brinkman equations.

\subsection{Strong dissipation}

In the opposite limit of very strong momentum dissipation,  $a\gg \lambda $, the integral equation can again be solved analytically. The Bessel function in (\ref{explicit-kernel}) is then exponetially small, and the kernel can be approximated by
\begin{equation}
 K(x)\stackrel{\lambda \rightarrow 0}{\simeq }\frac{1}{\pi x}+\ldots 
\end{equation}
The integral operator with this kernel does not have zero modes leading to an appatent contradiction. We need to recall that the flow profile develops two humps near the edges of the aperture and in the limit $\lambda \rightarrow 0$ the humps become infinitely high. Effectively this changes the boundary conditions allowing for more singular behavior at the edges, namely we need to allow for a $1/\sqrt{a^2-4x^2}$ asymptotics of the current profile. The solution with these more relaxed boundary conditions indeed exists:
\begin{equation}\label{flow-prof-lambda=0}
 J(x)\stackrel{\lambda \rightarrow 0}{\simeq }\frac{2}{\pi \sqrt{a^2-4x^2}}\,.
\end{equation}
The normalization factor is again fixed by (\ref{normalization-J}). An apparent divergence is not a real feature of the exact solution but an artifact of the approximation, in reality the current reaches a maximum of order $J_{\rm max}\sim 1/\sqrt{a\lambda }$ and then relaxes to zero as prescribed by the true boundary conditions (\ref{true-behavior-edges}). 

\subsection{Resistance}
\label{Large-L-resistance}

The resistance can be calculated from (\ref{int-representation-resistance}), (\ref{differential-resistance}). Setting $x_0=0$, dropping $r(x)$ and taking into account that $K(-x)=-K(x)$, we get
\begin{equation}\label{inf-res}
 R=\frac{2}{\sigma }\fint\limits_{-a/2}^{a/2} dx\, \widehat{K}(x)J(x),
\end{equation}
with
\begin{equation}\label{k-hat}
 \widehat{K}(x)=\fint\limits_{x}^{\infty }dx'\,K(x').
\end{equation}
And here we encounter a problem, for 
the last integral logarithmically diverges on the upper limit. This is not very surprising, an infinite sample necessarily has an infinite resistance. As mentioned briefly after eq.~(\ref{Greens}) this divergence is regulated by the finite sample's size. Mathematically speaking, the divergence reflects a breakdown of the large-volume approximation at $x\gtrsim L$. Using the exact $K(x)$ and $r(x)$ at large $x$ makes the integral finite and well-defined but leads to a weak, logarithmic dependnce of the resistance on the sample's size. A careful analysis is relegated to the appendix~\ref{regres:app}, here we adopt a simple practical workaround consisting  in introducing a rigid cutoff at some $x'_{\rm max}\sim L$. As shown in the appendix~\ref{regres:app}, this simple prescription reproduces the result of the rigorous systematic calculation provided the cutoff is set to $x'_{\rm max}=2L/\pi $. The resulting kernel is given by
\begin{equation}\label{explicit-k-hat}
 \widehat{K}(x)=-\frac{1}{\pi }\,\widehat{\mathcal{K}}\left(\frac{|x|}{\lambda }\right),
 ~~~
 \widehat{\mathcal{K}}(s)=\frac{1}{s^2}+\ln\frac{\pi \lambda s}{2L}
 +\frac{K_1(s)}{s}\,.
\end{equation}
These formulas express the resistance as a convolution of the flow profile at the aperture with the weight function $\widehat{K}(x)$.

In the Ohmic regime, $a\gg \lambda $, the flow profile is given by 
(\ref{flow-prof-lambda=0}). To compute the kernel in the same approximation we keep the leading logarithmic asymptotics in the weight function: 
\begin{equation}
 \widehat{K}(x)\stackrel{\lambda \rightarrow 0}{\simeq }
 -\frac{1}{\pi }\,\ln\frac{\pi  |x|}{2L}\,.
\end{equation}
Integration in (\ref{inf-res}) yields:
\begin{equation}\label{reslambda=0}
 R\stackrel{\lambda \rightarrow 0}{\simeq }\frac{2}{\pi \sigma }\,\ln\frac{8L}{\pi a}\equiv R_0,
\end{equation}
much smaller than the naive estimate (\ref{classR}). The resistance scales logarithmically with the sample size, not linearly as one may expect on general grounds.

In the opposite regime of weak dissipation, $a\ll \lambda $, the flow profile is approximately semi-circular. Taking $J(x)$ from  (\ref{LOlambda=inf}) and keeping only the leading term in the weight function,
\begin{equation}
 \widehat{K}(x)\stackrel{\lambda \rightarrow \infty }{\simeq }
 -\frac{2\lambda ^2}{\pi x^2}\,,
\end{equation}
we get for the resistance:
\begin{equation}\label{large-lambda-res}
 R\stackrel{\lambda \rightarrow \infty }{\simeq }
 \frac{32\lambda ^2}{\pi \sigma a^2}\,.
\end{equation}
When conductivity is traded for viscosity and carrier density using the kinetic theory relation (\ref{viscosity}), this formula  matches with the result found by Guo, Ilseven, Falkovich and Levitov \cite{Guo2017higher}:
\begin{equation}
 R\simeq \frac{32\eta }{\pi e^2n^2a^2}\,.
\end{equation}
As argued in  \cite{Guo2017higher} (see also sec.~\ref{estimates-of-resistance}) the $1/a^2$ scaling of the hydrodynamic resistance makes it superballistic for sufficiently small $a$.

It is equally easy to take into account the first dissipative correction. The corrected flow profile is given by (\ref{NLOlambda=inf}). Expanding the weight function (\ref{explicit-k-hat}) to the next (logarithmic) order we get:
\begin{equation}
 \widehat{K}(x){\simeq }-\frac{2\lambda ^2}{\pi x^2}
 -\frac{3}{2\pi }\ln\frac{|x|}{\lambda }
 +\frac{\ln 4+1-2\gamma  }{4\pi }\,,
\end{equation}
 where $\gamma =0.5772\ldots $ is the Euler constant. Here we carefully kept the constant term so this formula is accurate to $\mathcal{O}(x^2/\lambda ^2)$. Doing the integral we get:
\begin{equation}\label{Ranalytic}
 R{\simeq }\frac{32\lambda ^2}{\pi \sigma a^2}+\frac{1}{\pi \sigma }\,\ln\frac{\mathbbm{C}\lambda }{a}+R_0,
\end{equation}
where $R_0$ is the resistance in the Ohmic regime at $\lambda =0$, given by (\ref{reslambda=0}). The constant $\mathbbm{C}$ is numerically large 
\begin{equation}
 \mathbbm{C}=8\,{\rm e}\,^{\frac{5}{4}-\gamma }=15.68\ldots 
\end{equation}
As a result the nominally subleading log-term is numerically important for moderate values of $\lambda/a$.
Albeit the formula was derived for $\lambda \ll a$ it happens to describe the dependence on $\lambda$ for an almost entire range of parameters, failing only for very small dissipation length where the log-term starts blowing up.

\begin{figure}[bt]
    \centering
    \includegraphics[width=0.86\linewidth]{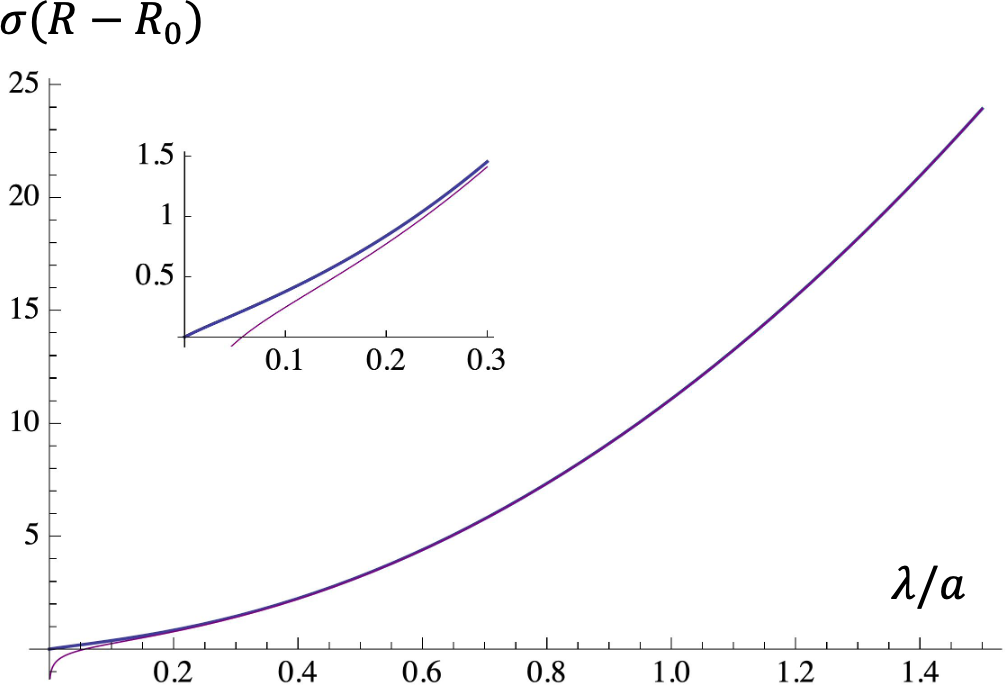}
    \caption{Resistance as a function of the dissipation length: the solid blue line is obtained by numerically solving the integral equation for $J(x)$. The thin purple line is the analytic approximation (\ref{Ranalytic}). The vertical axis is shifted to subtract the Ohmic resistance (\ref{reslambda=0}).}
    \label{fig:resistance-lambda}
\end{figure}

In fig.~\ref{fig:resistance-lambda} resistance  is plotted against the momentum dissipation length. The dependence on $\lambda $ can be viewed as a proxy for more easily measurable temperature dependence. Because $\lambda $ decreases with increasing temperature, the corresponding temperature scale is inverted, with higher temperatures corresponding to smaller values of $\lambda$.

Interestingly, our results predict a minimum of resistance at a certain temperature, a feature apparently observed in experiment \cite{Sarypov_2025}. For large temperatures, $\lambda $ goes to zero and the resistance approaches its asymptotic value (\ref{reslambda=0}). Its temperature dependence is then governed by the  bulk  conductivity and follows the usual flatiron law: $R\sim 1/\sigma \sim T$. At small temperatures, the dissipation length diverges as a power of $T$ and so does the conductivity. The resistance, given in that regime by (\ref{large-lambda-res}), depends more strongly on $\lambda $ and hence also diverges. Because resistance divereges at the two extremes there must be a minimum at some intermediate temperature.

\begin{figure}[bt]
    \centering
    \includegraphics[width=0.86\linewidth]{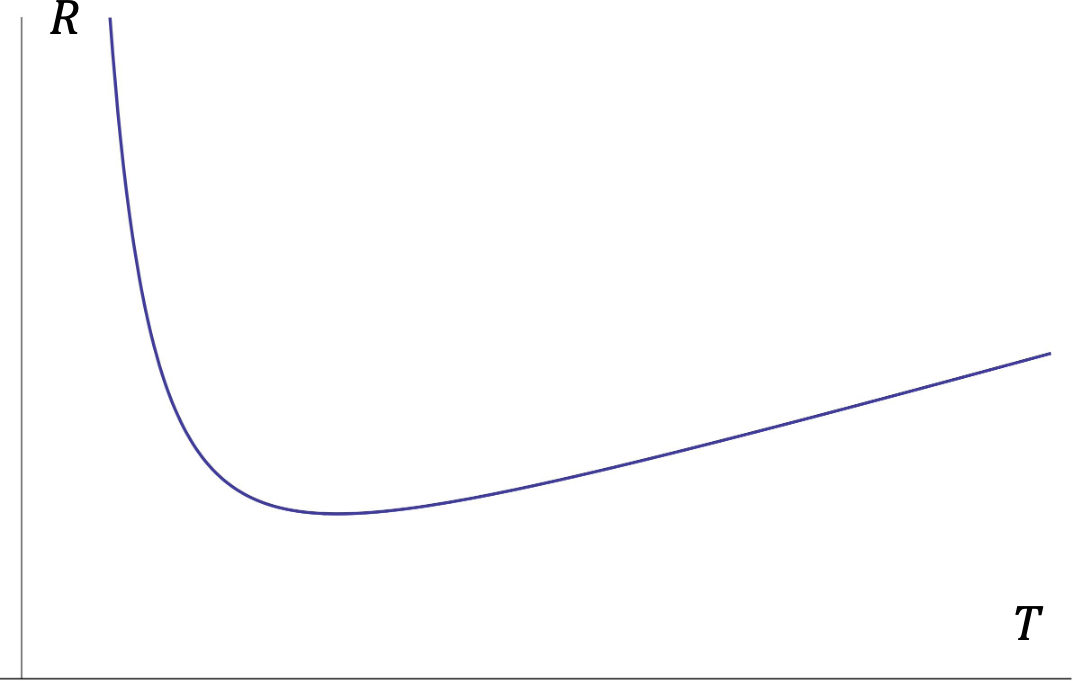}
    \caption{Resistance as a function of temperature. Both axes are plotted in arbitrary units.}
    \label{fig:resistance-T}
\end{figure}

We can sketch the temperature dependence of the resistance by adopting  the estimates of  \cite{2020PhRvB.102l5404P} for the electron-electron and impurity scattering length: $\ell_{\rm imp}\sim 1/T$, $\ell_{\rm ee}\sim 1/T^2$ (the "impurity" scattering  here rather refers to scattering on phonons  \cite{2020PhRvB.102l5404P}). Next, using the simple kinetic theory formulas for the momentum dissipation length and conductivity: $\lambda ^2\sim \ell_{\rm ee}\ell_{\rm imp}$, $\sigma \sim \ell_{\rm imp}$, we can estimate their temperature dependence as
\begin{equation}
 \lambda \sim T^{-3/2},\qquad \sigma \sim T^{-1}.
\end{equation}
With the help of these scaling laws we convert the $R-\lambda $ plot (fig.~\ref{fig:resistance-lambda}) into an $R-T$ plot  (fig.~\ref{fig:resistance-T}). At large $T$, $R\approx R_0\sim 1/\sigma \sim T$, while at small $T$, $R\sim \lambda ^2/\sigma \sim \ell_{\rm ee}\sim 1/T^2$, and indeed  at a certain temperature resistance has a minimum.

\section{Ohmic flow}

As another example where the problem can be solved fully analytically we consider the Ohmic flow with negligible dissipation length: 
$L,a\gg \lambda $. Setting $\lambda =0$ in the kernel (\ref{kernel-def}) and in the driving term (\ref{driving-term-def}) is equivalent to keeping the leading $\omega \rightarrow \infty $ asymptotics  of the integrands in the Fourier representation:
\begin{align}\label{OhmicK}
 &K(x)\stackrel{\lambda \rightarrow 0}{\simeq }\int\limits_{0}^{\infty }
 \frac{dk}{\pi }\,\,\coth kL\,\sin kx=\frac{1}{2L}\,\coth\frac{\pi x}{2L}
\nonumber \\
&r(x)\stackrel{\lambda \rightarrow 0}{\simeq }\int\limits_{0}^{\infty }
 \frac{dk}{\pi }\,\,\frac{\sin kx}{\sinh kL}=\frac{1}{2L}\,\tanh\frac{\pi x}{2L}\,.
\end{align}
The first integrand requires regularization and contains a divergent contribution which however does not contribute to the integral equation (\ref{main-integral-equation}) and has been simply dropped. An accurate analysis is carried out in the appendix~\ref{reg-Ohmic:sec}.

We are thus led to the following integral equation
\begin{equation}
 \fint\limits_{-a/2}^{a/2}dx'\,\coth\frac{\pi (x-x')}{2L}\,J(x')=\tanh\frac{\pi x}{2L}\,.
\end{equation}
This equation can be solved analytically. The solution with the strict square-root boundary conditions is actually unique and hence incompatible with normalization (\ref{normalization-J}). One extra parameter, needed for imposing the normalization condition, can be gained by relaxing the boundary conditions and allowing for the $1/\sqrt{a^2-4x^2}$ asymptotics at the edges. The ensuing endpoint singularity is a manifestation of the skin effect. The exact 
solution at finite dissipation length goes to zero at the endpoints 
but has two peaks distance $\lambda$ away from the edges, 
which in the limit of $\lambda\rightarrow 0$ become infinitely sharp 
and merge onto the boundary.  

Upon introducing the shorthand notations
\begin{align}
\label{change-of-vs}
 &\xi =\frac{\pi x}{2L}\,,
 \qquad 
 \alpha =\frac{\pi a}{4L}\,,
 \nonumber \\
 &f(\xi )=\frac{2L}{\pi }\,J\left(\frac{2L\xi }{\pi }\right),
 \qquad 
g(\xi )=\tanh \xi ,
\end{align}
the equations take the form:
\begin{align}\label{geninteq}
 &\fint\limits_{-\alpha }^{\alpha }d\xi '\,\coth(\xi -\xi ')f(\xi ')=g(\xi )
\\
\label{norma-f}
&\int\limits_{-\alpha }^{\alpha }d\xi \,f(\xi )=1.
\end{align}
Integral equations of this type can be reformulated as a Riemann-Hilbert problem, by intriducing a resolvent:
\begin{equation}\label{resolv}
 \widehat{\mathcal{R}}(\zeta )=\int\limits_{-\alpha }^{\alpha }d\xi '\,
 \coth(\zeta -\xi ')f(\xi '),
\end{equation}
a function of the complex variable $\zeta $ with the following properties:
\begin{itemize}
 \item The resolvent is $i\pi $-periodic: $\widehat{\mathcal{R}}(\zeta +i\pi n)=\widehat{\mathcal{R}}(\zeta )$ and thus naturally defined on a cylinder $\zeta \sim \zeta +i\pi $.
 \item It is analytic on the cylinder with a cut along the interval 
 $(-\alpha ,\alpha )$ of the real axis. On the cut:
\begin{equation}
 \widehat{\mathcal{R}}(\xi \pm i\varepsilon )=g(\xi )\mp i\pi f(\xi ).
\end{equation}
This condition is completely equivalent to the integral equation (\ref{geninteq}).
 \item The normalization condition (\ref{norma-f}) implies that at infinity the resolvent asymptotes to one:
\begin{equation}
 \widehat{\mathcal{R}}(\zeta )\stackrel{\mathop{\mathrm{Re}}\zeta \rightarrow +\infty }{\simeq }
 1+\mathcal{O}\left(\,{\rm e}\,^{-2\zeta }\right).
\end{equation}
\end{itemize}

These analytic data uniquely specifies $\widehat{\mathcal{R}}(\zeta )$ as a function of a complex variable. Reconstring a function from an analytic data is known as the Riemann-Hilbert problem, in this case it is solved by the following integral transform \cite{Gakhov}:
\begin{align}
 \widehat{\mathcal{R}}(\zeta )=&\int\limits_{-\alpha }^{\alpha }
 \frac{d\eta \,g(\eta )}{\pi \sinh(\zeta -\eta )}\,\,
 \sqrt{\frac{\cosh^2\zeta -\cosh^2\alpha }{\cosh^2\alpha-\cosh^2\eta }}
\nonumber \\
&+\frac{A\cosh\zeta }{\sqrt{\cosh^2\zeta -\cosh^2\alpha }}\,.
\end{align}
The constant $A$ is fixed by the asymptotics at infinity:
\begin{equation}
 A=1-\int\limits_{-\alpha }^{\alpha }
 \frac{d\eta \,g(\eta )}{\sqrt{\cosh^2\alpha  -\cosh^2\eta  }}\,.
\end{equation}

For $g(\eta )=\tanh \eta $ the integral can be computed explicitly:
\begin{align}\label{resolvent-in-condensed-f}
 \widehat{\mathcal{R}}(\zeta )=&\tanh\zeta -\frac{\sqrt{\cosh^2\zeta -\cosh^2\alpha }}{\cosh\zeta \cosh\alpha }
\nonumber \\
 &+\frac{\cosh\zeta }{\cosh\alpha \sqrt{\cosh^2\zeta -\cosh^2\alpha }}\,.
\end{align}
The imaginary part of the resolvent (basically the last two terms) is the unknown function $f(\xi )$. Returning to the original notations we get for the flow profile:
\begin{align}
 J(x)=&\frac{1}{2L\cosh\frac{\pi a}{4L}}
 \left(
 \frac{\sqrt{\cosh^2\frac{\pi a}{4L}-\cosh^2\frac{\pi x}{2L}}}
 {\cosh\frac{\pi x}{2L}}
 \right.
\nonumber \\
&\left.
 + \frac {\cosh\frac{\pi x}{2L}}
 {\sqrt{\cosh^2\frac{\pi a}{4L}-\cosh^2\frac{\pi x}{2L}}}
 \right).
\end{align}

\begin{figure}[bt]
    \centering
    \includegraphics[width=0.9\linewidth]{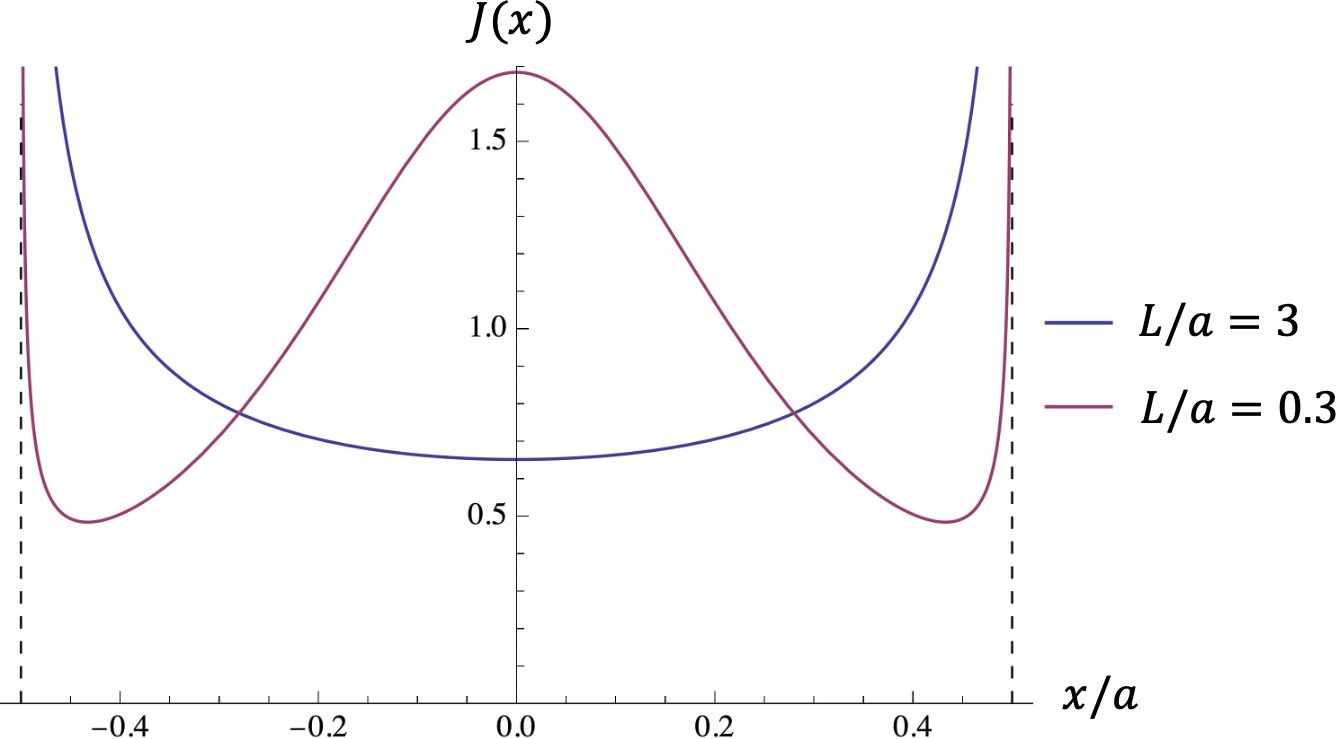}
    \caption{The Ohmic flow.}
    \label{fig:OhmicFl}
\end{figure}

For sufficiently large $L$ the morphology of the flow is the same as in the infinite-size limit: the current has two peaks at the edges of the constriction and a trough in the middle. The peaks, infinitely high in this approximation, represent the skin effect. With lowering $L$ a third maximum develops in the middle of the aperture. The peaks at the edges are then followed by the dips with the current growing towards the centre of the aperture. This is illustrated in Fig.~\ref{fig:OhmicFl}.

The resistance can be calculated from (\ref{int-representation-resistance}), (\ref{differential-resistance}). We notice first that the differential resistance (\ref{differential-resistance}) is related to the resolvent introduced in (\ref{resolv}). Taking into account (\ref{OhmicK}) and the change of variables (\ref{change-of-vs}), we find:
\begin{equation}
 \mathcal{R}(x)=\frac{1}{2L}\left(\widehat{\mathcal{R}}\left(\frac{\pi x}{2L}\right)-\tanh \frac{\pi x}{2L}\right).
\end{equation}
Choosing $x_0=a/2$ in (\ref{int-representation-resistance}) we get:
\begin{equation}
 R=\frac{2}{\pi \sigma }\int\limits_{\alpha }^{\infty  }
 d\xi \,\left(\widehat{\mathcal{R}}(\xi )-\tanh\xi \right).
\end{equation}
Interestingly, the integrand is basically the analytic continuation of the flow profile past the branch point at the edge of the aperture. Using the explicit solution,
\begin{align}
 R=\frac{2}{\pi \sigma \cosh\alpha }\int\limits_{\alpha }^{\infty  }
 d\xi \,&\left(
 \frac{\cosh\xi }{ \sqrt{\cosh^2\xi -\cosh^2\alpha }}
 \right.
\nonumber \\
 &\left. - \frac{\sqrt{\cosh^2\xi -\cosh^2\alpha }}{\cosh\xi}
 \right),
\end{align}
 and the resistance integrates to
\begin{equation}
 R=\frac{1}{\pi \sigma }\,\ln\frac{\cosh\frac{\pi a}{4L}+1}{\cosh\frac{\pi a}{4L}-1}\,.
\end{equation}
For $L\gg a$:
\begin{equation}
 R\stackrel{L\gg a}{\simeq }\frac{2}{\pi \sigma }\,\ln\frac{8L}{\pi a}\,,
\end{equation}
which matches with (\ref{reslambda=0}), while for $L\ll a$ the resistance is exponentially small:
\begin{equation}
 R\stackrel{L\ll a}{\simeq }\frac{4}{\pi \sigma }\,\,{\rm e}\,^{-\frac{\pi a}{4L}}\,,
\end{equation}
and also here the flow can be superballistic, the constriction with a big aperture does not impede the current too much.

\section{No constriction}

The regime of  $a\gg L,\lambda $ corresponds to removing the constriction: $a\rightarrow \infty $. The integral equation (\ref{main-integral-equation}) then takes on a convolution form and is solved by Fourier transform:
\begin{widetext}
\begin{equation}
 J(x)=2\int\limits_{0}^{\infty }\frac{dk}{\pi}\,\,
 \frac{\omega \sinh\omega -k\sinh k}{(\omega -k)\sinh(\omega +k)L+(\omega +k)\sinh(\omega -k)L}\,\cos kx. 
\end{equation}
\end{widetext}
The current is streamed into a jet with the flow quickly  (exponentially) decaying outside of it, and 
the flow profile has a lump shape, concentrated within a region of width $w$.
The width of the jet can be estimated as
\begin{equation}
 w^2=\int_{}^{}dx\, x^2J(x)=\frac{L^2}{2}\left(1-\frac{2\lambda }{L}\,\tanh\frac{L}{2\lambda }\right),
\end{equation}
 and is basically controlled by the size of the sample, but can be much smaller if dissipation is low and $\lambda \gg L$.

\section{Numerical simulations}

For more complex boundary conditions the Brinkman Eq.~(\ref{eq:brinkman}) can be solved numerically using the finite-element method (FEM). As an example we study the flow through
a channel of finite width that interpolates between the  flow through
an infinitely long channel and the geometry we studied so far. Specifically we want to verify if the skin effect survives in a more complicated geometry.
In the FEM formulation, the system is modeled as a flow through a pipe with a constriction in a form of perpendicular wall with a channel of a given width $a$ and length $l$. On the boundaries $\Gamma$ of the pipe and the constriction wall, we impose no-slip boundary conditions: $j_i(\Gamma)=0$, or, in Sec.~\ref{maxwellbc}, replace them with Maxwell partial-slip boundary conditions.
At the inlet and outlet of the pipe we impose \emph{Poiseuille flow} conditions.
Along the aperture (at the center of the channel), we impose a constant potential ($\varphi_0$), consistent with the assumptions made for the analytical considerations. A detailed description of the numerical method and the geometry of the computational domain (box) can be found in the Appendix~C.
\begin{figure}[bt]
    \centering
    \includegraphics[width=.95\linewidth]{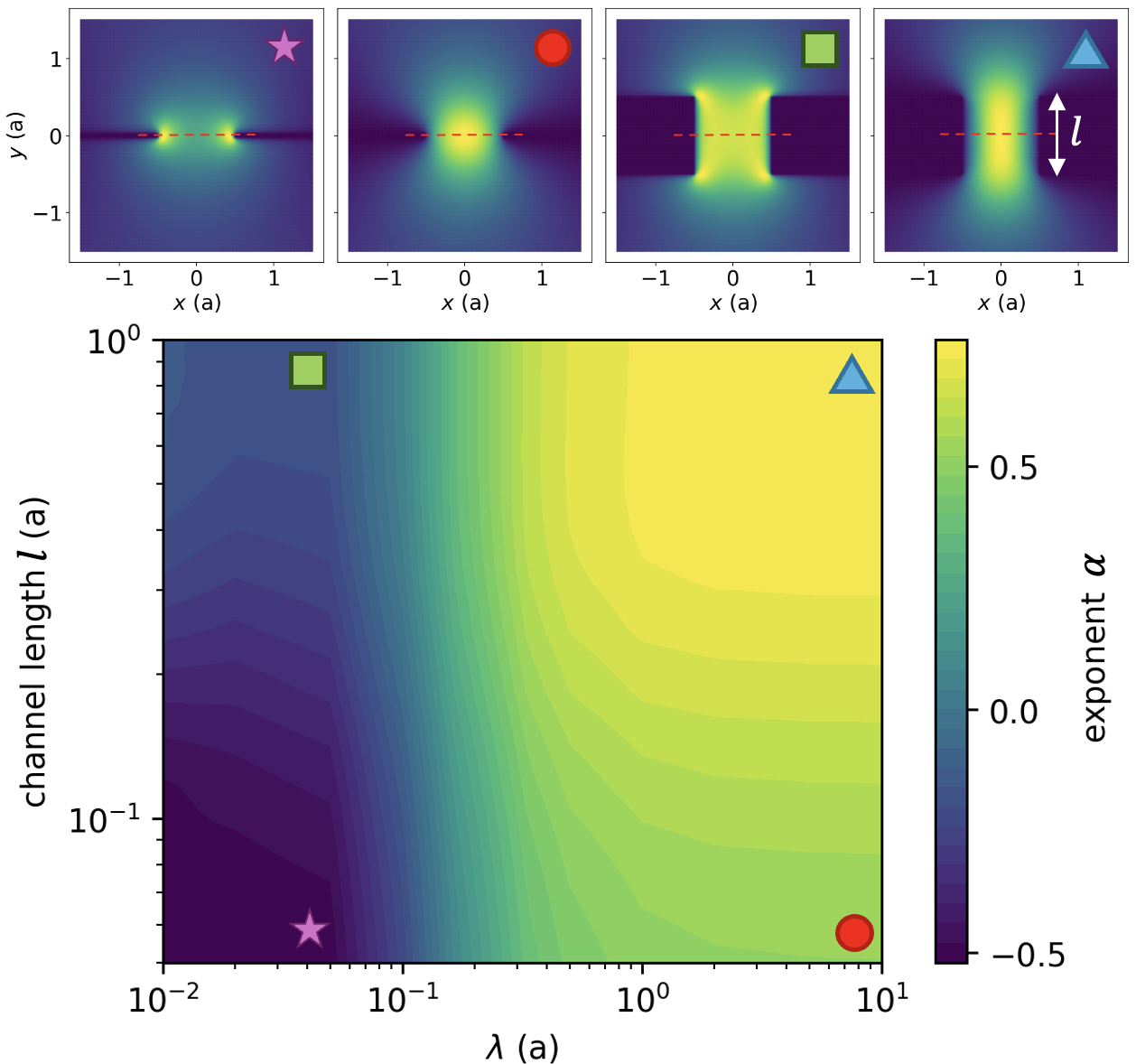}
    \caption{FEM results for the flow through the channel with a finite length $l$: maps showing the exponent $\alpha$ fitted to the velocity profile $J$ across the aperture (the fitting line is marked by red dashed lines in the upper panels). Representative cases, with parameter $(\lambda,l)$ configurations -- showing four different flow regimes -- indicated by symbols on the map, are shown in the upper panels.}
    \label{fig:fit_exp}
\end{figure}

\subsection{Finite channel length}
FEM calculations for the shallow channel of length $l=a/20$ yield very good agreement between the velocity profiles and the analytical results given by Eqs.~(\ref{Fouriervx}) and (\ref{Fouriervy}). A detailed comparison can be found in Appendix~C, confirming the validity of the assumptions made in the analytical considerations.
We also extend our results to more realistic geometries with finite-length channels, as presented in Fig.~\ref{fig:fit_exp}.

Fitting the current $j_y(x)$ profiles across the channel aperture to the form $A\!\left(a^2-4x^2\right)^\alpha$ yields a map of the exponent $\alpha$, revealing the crossover from the Poiseuille-like profile (with $\alpha>0.7$ marked by triangle) through the semicircular profile (circle, $\alpha\sim 0.5$), Eq.~(\ref{LOlambda=inf}), to the skin-effect profile (star, $\alpha\sim -0.5$).
Detailed profile plots can be found in Appendix~D.
Phase diagram containing viscous, ballistic, and Ohmic regimes of the current profile at a constriction was also discussed in~\cite{Huang:2021mee}.
The Brinkman formulation provides a minimal and analytically tractable description of the 
viscous-to-Ohmic crossover in constricted geometries, which can be further numerically extended to channels of finite length.
The parameter $\lambda$ controls the spatial decay of the flow and encodes momentum-relaxing processes. In the low dissipation regime $\lambda \gg a$, 
the current obeys the Stokes flow with 
the characteristic single-peak Poiseuille-like (long channel) or semicircular profile (narrow channel), whereas for $\lambda \ll a$ the flow approaches the Ohmic limit. 

For intermediate and small values of $\lambda / a$ the flow features a skin effect thereby giving rise to a double-hump structure for the simple single-slit geometry. Such a double-hump profile has been observed experimentally in graphene~\cite{jenkins2022imaging}, and our semi-analytic solution shows reasonable agreement with the experimental data for $\lambda\simeq a/10$. Numerical simulation of more complex geometries suggest that the skin effect is a universal feature of viscous flows with momentum relaxation, emerging whenever the geometry contains sharp edges. An edge locally enhances the current density, leading to a relative suppression in regions where the flow appears otherwise unobstructed.

\subsection{Convection}
We also studied nonlinear effects by adding convective term $\sigma D \,j_k\partial_kj_i$ to the RHS of the Brinkman Eq.~(\ref{eq:brinkman}) in the FEM formulation.
To make convection observable in simulations, we assumed coefficient $D=D_0=10^{-6}\ \mathrm{m^2 V/A^2}$ such that for $\lambda = a/10$ convective and Ohmic terms in the aperture area are comparable (see Appendix~E for a detailed discussion). 
\begin{figure}[bt]
    \centering
    \includegraphics[width=0.99\linewidth]{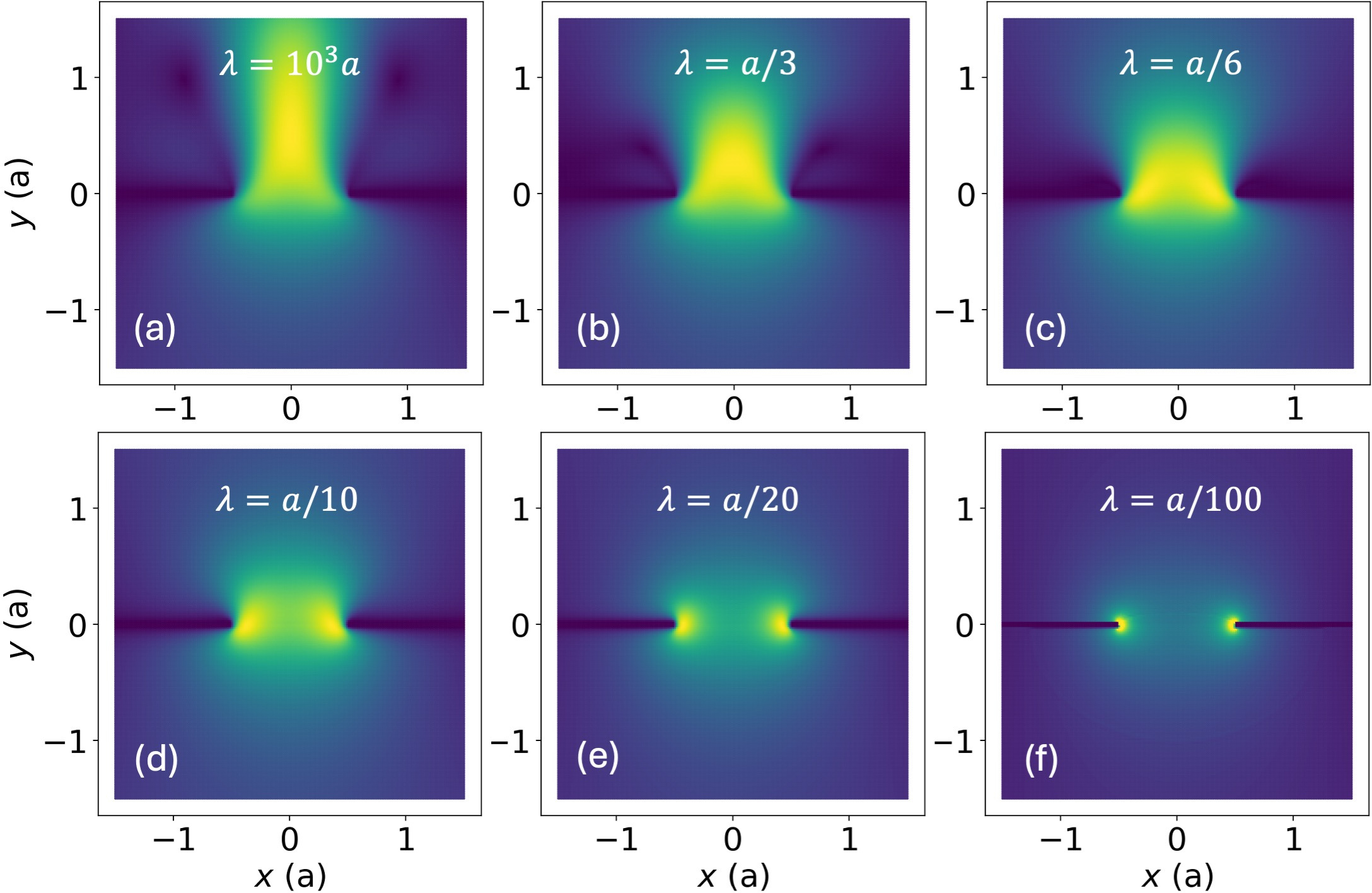}
    \caption{The current density $|j|$ for various momentum relaxation lengths (a-f): $\lambda=10^3a$, $a/3$, $a/6$, $a/10$, $a/20$, and $a/100$; with a convective term included in the FEM formulation.}
    \label{fig:fem_conv}
\end{figure}
One should note, however, that for such large values of $D$ may significantly overestimate convection effects in graphene (by more than a factor of 20, as estimated in Appendix~E). 
Results including the convective term are presented in Fig.~\ref{fig:fem_conv}. They exhibit two characteristic features: (1) the velocity profiles along the aperture, $j_y(y)$, become asymmetric, with the asymmetry reflecting the flow direction; and (2) a characteristic pair of vortices in the $\bm{j}$ flow emerges in the opening region for slow relaxation (large $\lambda$). These are kind of \emph{corner eddies} that emerge due to the presence of neighboring walls of the computational domain, perpendicular to the constriction, with no-slip boundary conditions. 

\subsection{Maxwell boundary conditions}
\label{maxwellbc}
\begin{figure}[b]
    \centering
    \includegraphics[width=0.99\linewidth]{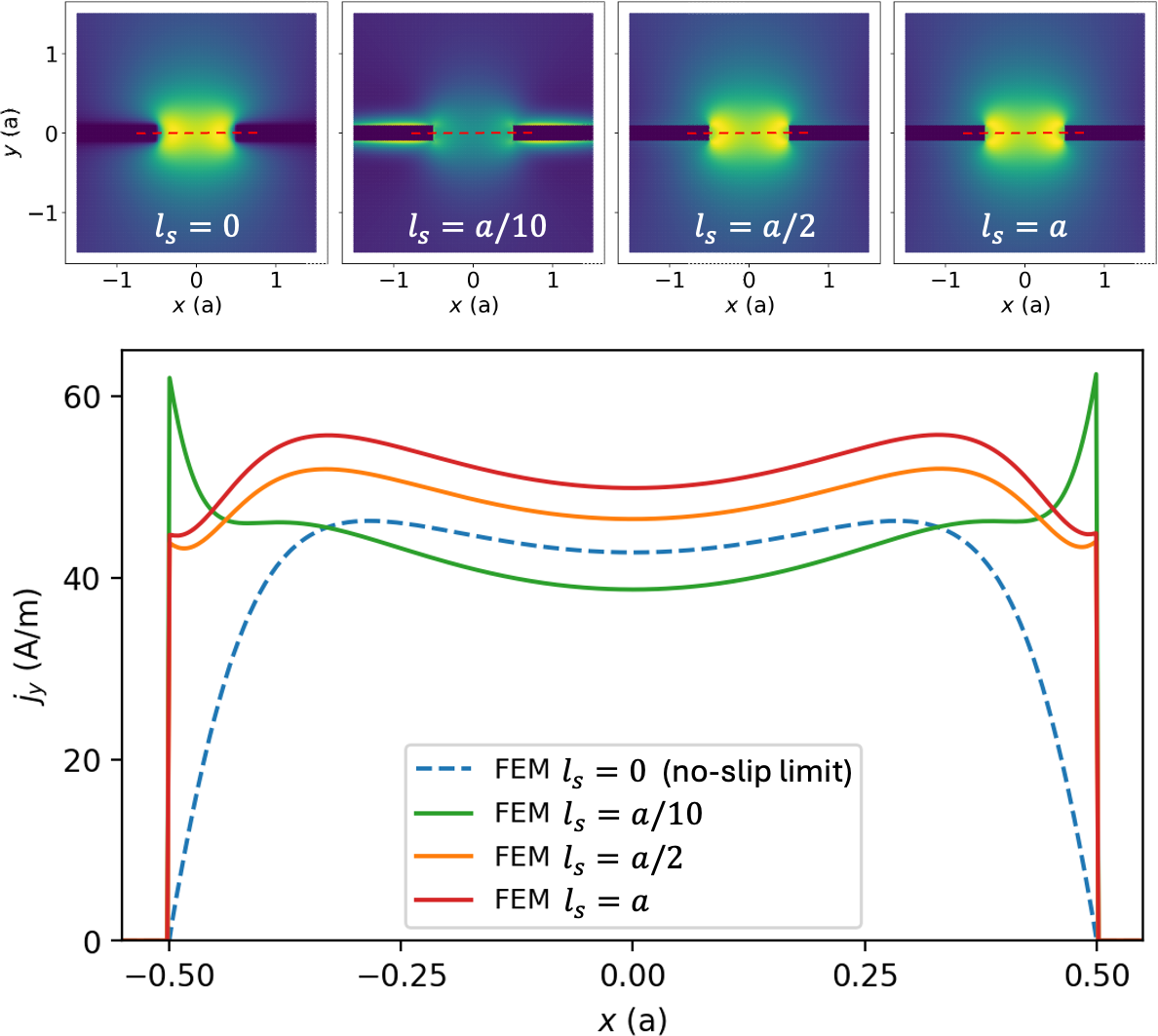}
    \caption{FEM results obtained using the Maxwell (partial-slip) boundary condition for different slip lengths $l_s$, with fixed momentum-relaxation length $\lambda=a/10$ and channel length $l=a/5$. Upper panels: current-density maps $|j|$ for $l_s=0$, $a/10$, $a/2$, and $a$, respectively. The red dashed line indicates the cut across the aperture used to extract the current profiles $j_y(x)$ shown in the lower panel.}
    \label{fig:max}
\end{figure}
To examine the role of boundary conditions, we replace the no-slip condition on the solid boundary $\Gamma$ by the impermeability and Maxwell partial-slip conditions \begin{equation} j_{\perp}\big|_{\Gamma}=0, \qquad j_{\parallel}\big|_{\Gamma} =l_s\,\partial_{r_{\perp}}j_{\parallel}\big|_{\Gamma}. \end{equation} Here, $\hat{\boldsymbol n}$ and $\hat{\boldsymbol t}$ are the unit normal and tangent to $\Gamma$, respectively, $j_{\perp}=\boldsymbol{j}\cdot\hat{\boldsymbol n}$ and $j_{\parallel}=\boldsymbol{j}\cdot\hat{\boldsymbol t}$, while $\partial_{r_{\perp}}\equiv\hat{\boldsymbol n}_{\mathrm{in}}\cdot\nabla$ denotes differentiation in the direction normal to the boundary and pointing into the fluid. The parameter $l_s$ is the slip length: $l_s=0$ gives the no-slip limit, whereas increasing $l_s$ permits a nonzero tangential current at the wall and approaches the stress-free limit.
This provides a continuous interpolation between strongly diffusive and partially specular boundary scattering, in the spirit of the Maxwell boundary condition discussed in Ref.~\cite{moessner_boundary-condition_2019}. In the finite computational box we also impose a Poiseuille inflow and outflow profile modified by the same slip length~\cite{moessner_boundary-condition_2019}: 
$j_y(\mathrm{in/out})=j_0(\tilde{L}_x^2-4x^2)/\tilde{L}_x^2$, with $\tilde{L}_x=L_x+2\tilde{l}_s=\sqrt{L_x^2+4L_xl_s}$, ensuring consistency between the wall boundary condition and the imposed current profile at the inlet and outlet. 
$\tilde{L}_x$ means \emph{extended} box on which the parabolic profile is expected. The resulting current-density maps and aperture profiles are shown in Fig.~\ref{fig:max}. For $l_s=0$, we have the reference double-bump profile as observed for no-slip boundaries. Importantly, the skin-effect-like redistribution of current persists for all finite values of $l_s$ considered, while increasing $l_s$ mainly modifies the quantitative shape of the aperture profile, making the double-hump structure even more pronounced. For small but finite slip lengths (green curve), a characteristic enhancement of the current density is observed near the slit edges. This enhancement reflects the large shear required to satisfy the Maxwell boundary condition when the wall velocity becomes finite, resulting in a locally increased current density.

These results demonstrate that, within the Brinkman model and for the finite slip lengths investigated here, the edge-enhanced current distribution is not an artifact of imposing a no-slip boundary condition. Its persistence under partial slip instead reflects the combined influence of constriction geometry and momentum relaxation.

\section{Discussion}

Our results offer a complementary interpretation of current flow profiles observed in constricted graphene geometries, such as those imaged by scanning NV magnetometry. In the earlier analyses based on kinetic theory \cite{Guo2017higher,heilmann2024transition,reingruber2025quantum,Starkov2025}, double-peaked current profiles were commonly associated with the Ohmic regime dominated by momentum-relaxing scattering, while single-peaked profiles were taken as signatures of ballistic or weakly hydrodynamic transport. Within the Brinkman hydrodynamic framework adopted here, this identification is not unique. We show that edge-enhanced, double-hump profiles arise naturally in a fully hydrodynamic description once finite momentum relaxation is included, as a consequence of a geometry-induced skin effect. In this picture, within
the Brinkman framework and for the constriction geometries considered here, the redistribution of
current toward sharp edges can occur in a viscous flow with finite momentum relaxation. Consequently, an edge-enhanced profile alone does not uniquely identify a non-hydrodynamic transport
regime. The constriction geometry thus plays an active role in shaping the flow, and similar profiles can emerge across a broad range of hydrodynamic parameters. Our analysis highlights that current profile shapes alone do not uniquely determine the underlying transport regime, and that geometric effects must be carefully disentangled from microscopic scattering mechanisms when interpreting experiments in constricted electronic systems in anlogy with the AC effects \cite{moessner_pulsating_2018,moessner_boundary-condition_2019,cosme_electronic_2022}.

Finally, we note that the double-hump profile also arises in a purely viscous fluid under more complex boundary conditions~\cite{Asafov2022}, where a finite slip length plays a role analogous to momentum dissipation in our approach.

\acknowledgments
We would like to thank C.~Arratya, N.~Ben-Schachar, M.~Chernodub, G.~Falkovich and D.~Mitra for very interesting discussions. We also thank M.~Chernodub and V.~Goy for collaboration on the early stages of this project. P.S. was supported in part by the Polish National Science Centre (NCN) Sonata Bis grant 2019/34/E/ST3/00405. The work of K.Z. was supported by VR grant 2021-04578.


\appendix
\section*{Appendices}

\appendix

\section{Large-volume resistance}
\label{regres:app}

In sec.~\ref{Large-L-resistance} we encountered a divergent integral in calculating the resistance of an infinite sample. Since the divergence is caused by the far-away region, the integral (\ref{int-representation-resistance}) can be calculated by splitting the domain of integration in two parts:
\begin{equation}
 R=R_{c}+\Delta R,
\end{equation}
where $R_c$ is the contribution from $x_0<x<L_c$ and $\Delta R$ accounts for $x>L_c$.
 Choosing the matching scale at $L_c\gg \lambda,a $, but $L_c\ll L$ we can use all the approximations of sec.~\ref{sec:large-L} for $x<L_c$ and for $x>L_c$ we can neglect $x'$ compared to $x$ in the integral (\ref{differential-resistance}), so the far-away contribution is simply
\begin{equation}\label{rema}
 \Delta R=\frac{2}{\sigma }\int\limits_{L_c}^{\infty }dx\left(K(x)-r(x)\right).
\end{equation}

The result for the finite part is essentially (\ref{inf-res}),  (\ref{explicit-k-hat}) but with the cutoff set to $L_c$ rather than $2L/\pi $:
\begin{equation}\label{app-explicit-k-hat}
 \widehat{K}_c(x)=-\frac{1}{\pi }\,\widehat{\mathcal{K}}_c\left(\frac{|x|}{\lambda }\right),
 ~~~
 \widehat{\mathcal{K}}_c(s)=\frac{1}{s^2}+\ln\frac{\lambda s}{L_c}
 +\frac{K_1(s)}{s}\,.
\end{equation}
We will see that the only effect of the far-away contribution is to shift the cutoff from $L_c$ to $2L/\pi $.

To calculate the remainder we notice that both $x$ and $L$ in the integral (\ref{rema}) are much larger than $\lambda$ and we can set $\lambda=0$ when calculating   
the functions $K(x)$ and $r(x)$. Operationally this amounts to setting $\omega=\infty$ in the Fourier integrals (\ref{kernel-def}), (\ref{driving-term-def}), then:
\begin{equation}
    \Delta R=\frac{2}{\sigma}\int\limits_{L_c}^{\infty}dx\,
    \int\limits_{0}^{\infty}\frac{dk}{\pi}\,\,
    \tanh\frac{kL}{2}\,\sin kx.
\end{equation}
The inner integral is also not terribly well defined. As written it does not converge well at infinity, which can be remedied via integration by parts. in a fully rigorous calculation integration by parts should be the very first step, and should be done before any approximations are made, the boundary terms would then automatically cancel. We just assume no boundary terms contribute:
\begin{equation}
    \Delta R=\frac{L}{\sigma }\int\limits_{L_c}^{\infty}\frac{dx}{x}\,\,
    \int\limits_{0}^{\infty}\frac{dk}{\pi}\,\,
    \frac{\cos kx}{\cosh^2\frac{kL}{2}}\,.
\end{equation}
The integrals can be now computed without troubles and for $L\gg L_c$ give:
\begin{equation}
    \Delta R=\frac{2}{\pi\sigma}\,\ln\frac{2L}{\pi L_c}\,.
\end{equation}
When added to the finite part written in the form (\ref{inf-res}) 
with the kernel 
(\ref{app-explicit-k-hat}), the remainder cancels the dependence on $L_c$ as it should, and effectively replaces the matching scale by the physical parameter $2L/\pi$. 

\section{Regularized kernel in Ohmic regime}\label{reg-Ohmic:sec}

The Fourier integral for the kernel of the integral equation in the Ohmic ($\lambda \rightarrow 0$) approximation  (\ref{OhmicK}) diverges at large $k$. The limits $\lambda \rightarrow 0$ and $k\rightarrow \infty $ actually do not commute, and Fourier integral for the original, finite-$\lambda $ kernel is fully convergent. The Ohmic approximation breaks down at large wave numbers, where setting $\lambda =0$ can no longer be trusted. To regularize the integral we simply remove large wave numbers by imposing a hard cutoff at $k=\Lambda_c\sim 1/\lambda $:
\begin{equation}
 K_{\rm reg}(x)=\int\limits_{0}^{\Lambda _c}
 \frac{dk}{\pi }\,\,\coth kL\,\sin kx,
\end{equation}
which can be written as
\begin{equation}
 K_{\rm reg}(x)=\int\limits_{0}^{\infty }
 \frac{dk}{\pi }\,\,(\coth kL-1)\sin kx
 +\int\limits_{0}^{\Lambda _c}
 \frac{dk}{\pi }\,\,\sin kx,
\end{equation}
where the first term is finite does not need regularization. Doing the integrals we get:
\begin{equation}
 K_{\rm reg}(x)=\frac{1}{2L}\,\coth\frac{\pi x}{2L}-\frac{\cos \Lambda _c x}{\pi x}\,.
\end{equation}
The second term is in principle there, and becomes important for $x\sim 1/\Lambda _c$ removing the $1/\pi x$ singularity from the finite part. But this is just equivalent to the principal value prescription used throughout all our calculations. The extra term wildly oscillates and its convolution against any smooth, slowly-varying function would go to zero when the cutoff is removed. We have simply dropped this term when analyzing the integral equation in the main text.

\section{Finite element calculations in finite domain}

\begin{figure}[bt]
    \centering
    \hspace{4mm}
    \includegraphics[width=0.85\linewidth]{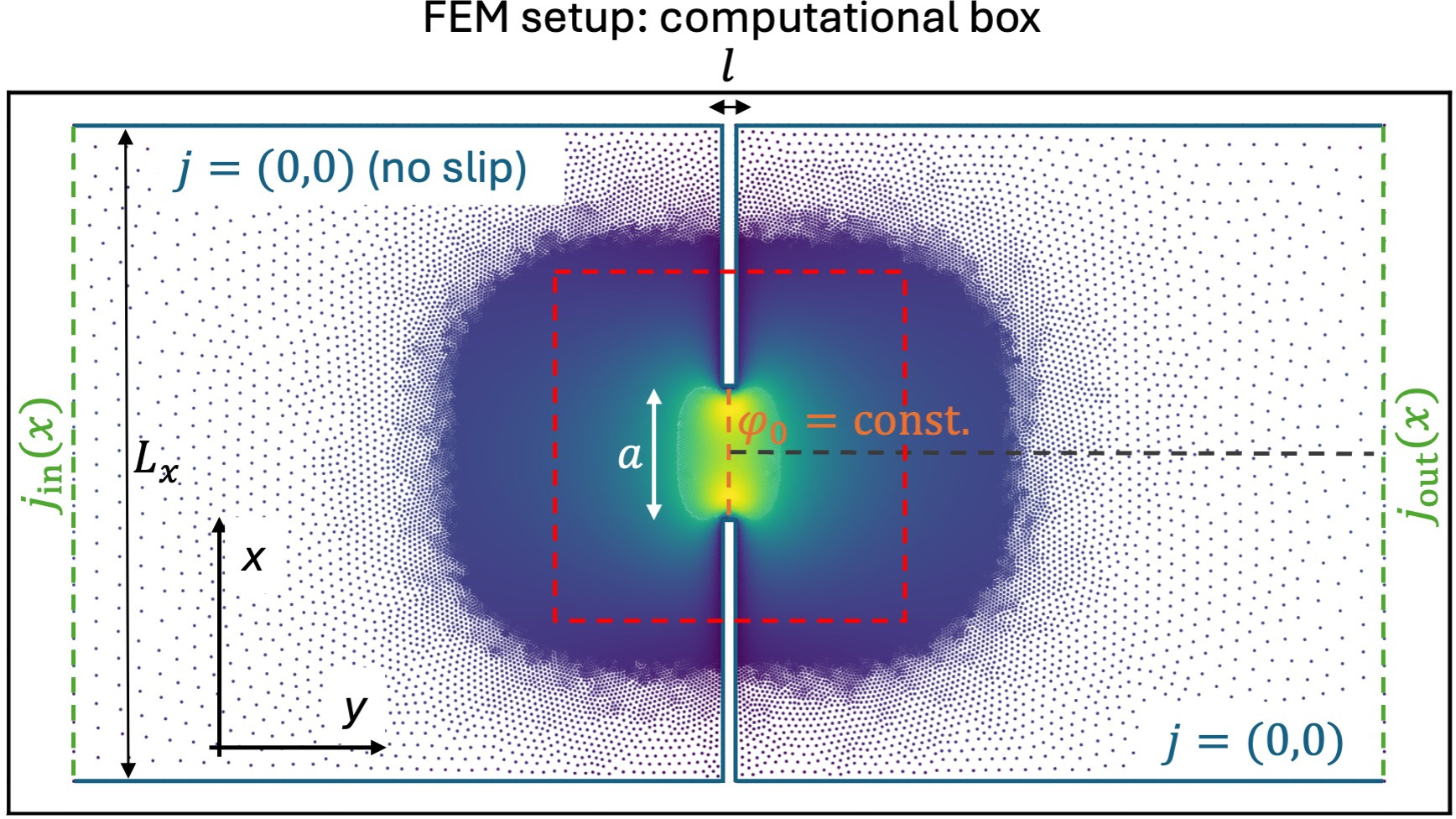}\\
    \vspace{2mm}
    \includegraphics[width=0.99\linewidth]{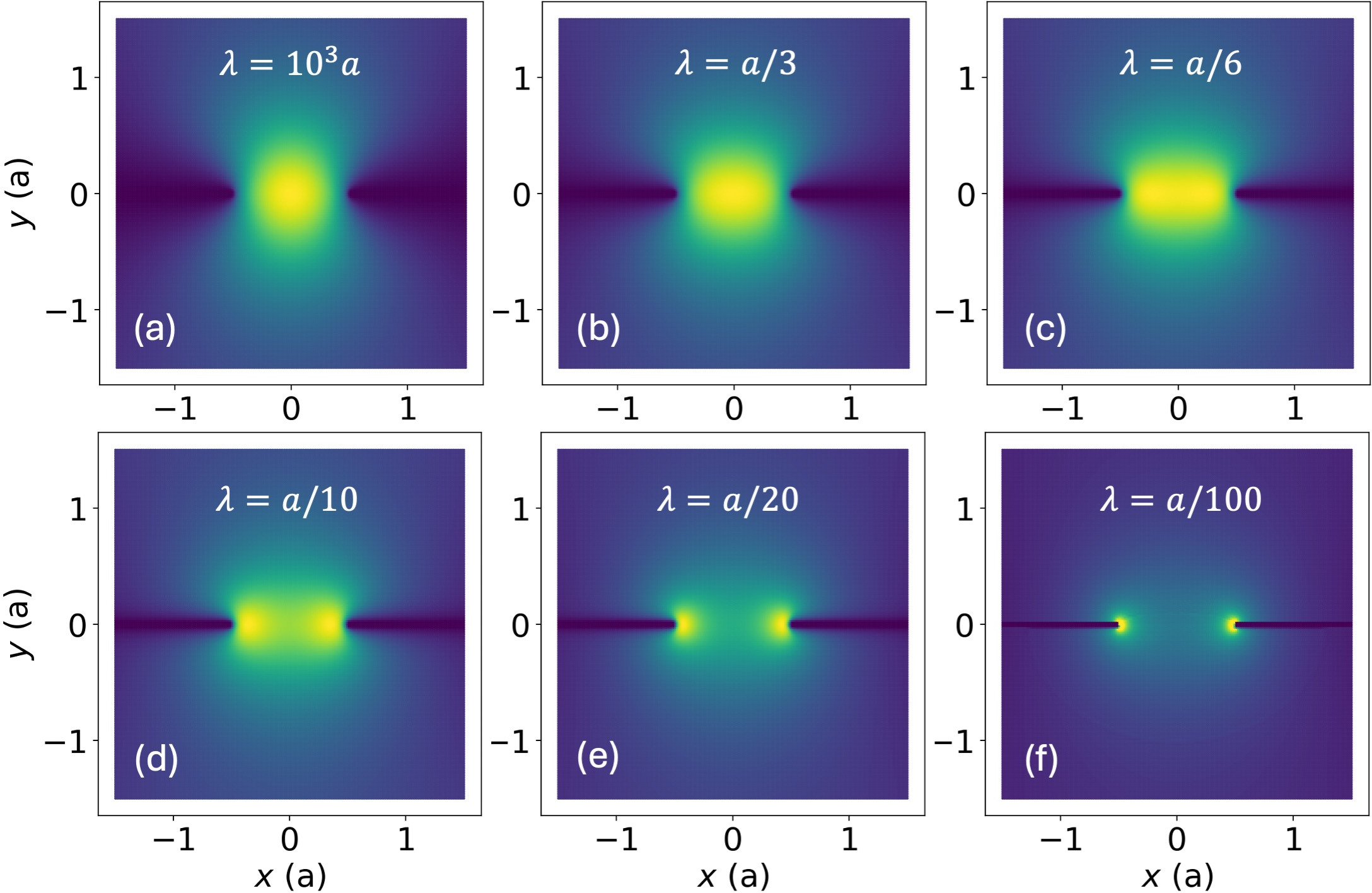}
    \caption{(top) FEM computational box with marked various boundary conditions. (bottom) FEM calculations showing current density $|j|$ for various momentum relaxation length (a-f): $\lambda=10^3\,a$, $a/3$, $a/6$, $a/10$, $a/20$, and $a/100$. Cf. analytical results in Fig.~\ref{fig:an}.}
    \label{fig:FEM}
\end{figure}
\begin{figure*}[tb]
    \centering
    \includegraphics[width=.45\linewidth]{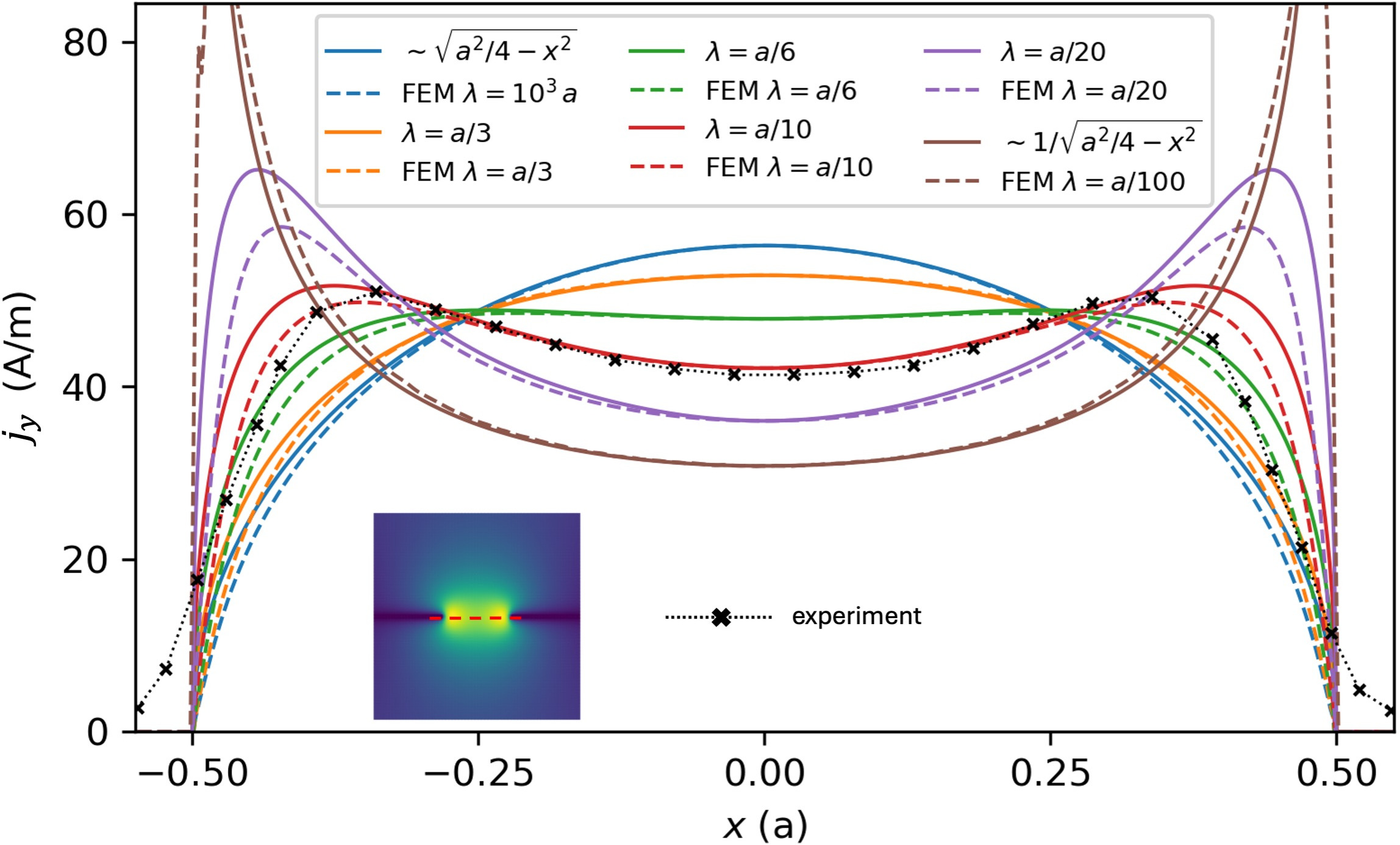}
    \hspace{2mm}
    \includegraphics[width=.35\linewidth]{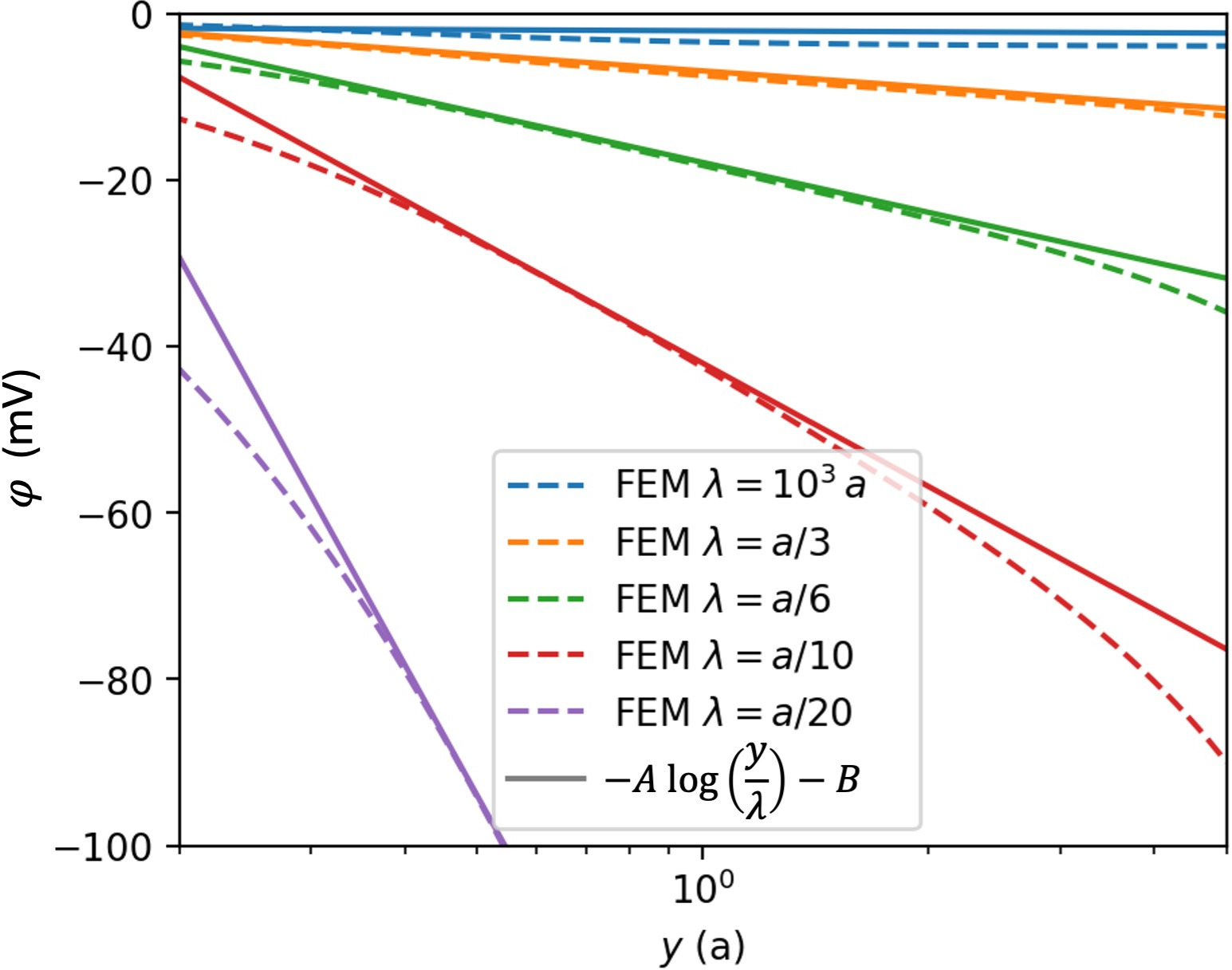}
    \caption{
    (left) The longitundinal current $j_y\equiv J$ at midflow (cut section shown in the inset as a red dashed line) comparing analytical solution of the Brinkman equations for an infinitely thin constriction (solid lines) with the numerical FEM solution for the channel of finite length $l=a/20$ (dashed lines). Experimental data (dotted black line) is taken from \cite{jenkins2022imaging}. (right) Electrostatic potential profiles across the device.}
    \label{fig:fem_vs_an}
\end{figure*}

To get numerical solutions of the modeled system we use the finite element method (FEM) with a proper variational formulation that gives discretization of the Brinkman equations~(\ref{eq:brinkman}) on the grid that is adapted to the geometry of our problem. In the presented calculations we have used a simple but efficient \textit{splitting method} also known as \textit{Chorin method}~\cite{Chorin1968} or \textit{incremental pressure correction scheme} (IPCS)~\cite{Goda1979}.
To implement the FEM calculations we have used FEniCS library~\cite{Alnaes2015,Logg2010} which enables convenient expression of equations in their weak formulation through the UFL language~\cite{Alnaes2014}. Meshes were created using the Gmsh library~\cite{Geuzaine2009}.

Let us also comment on two important differences between the problem definition in analytical and numerical domains. Due to the fact that in numerics we have to deal with a finite area, i.e. a computational box that encloses the channel surrounding, in contrast to analytical domain where we solved the equations for infinite half-plane, our numerical formulation resembles rather \textit{Poiseuille flow} through a rectangular pipe with a constriction in a form of a perpendicular wall with a channel of width $a$ and length $l$. To define a problem in a finite area we have to set up the proper conditions on the computational box boundary as presented in Fig.~\ref{fig:FEM}(top). 
At the inlet (left side) and outlet (right side) of the pipe (marked by dashed green lines) we impose a Poiseuille flow: $\bm{j}_\mathrm{in}(x)=\bm{j}_\mathrm{out}(x)=(0,j_0(L_x^2-4x^2)/L_x^2)$. At the top and bottom edges and on the constriction wall $\Gamma$ (marked by the blue solid line) no-slip boundary conditions $j_i(\Gamma)=0$ are enforced.
At the center of the channel -- across the aperture (marked by the orange dashed line), we impose a constant potential 
$\varphi_0$. In the vicinity of the channel, the mesh is appropriately refined to improve numerical accuracy.
Fig.~\ref{fig:FEM}(bottom) displays the current flow in the zoomed-in region marked by the red dashed square in Fig~\ref{fig:FEM}(top). The results are similar to the analytic solution for the flow in the infinite domain presented in the main text (c.f. Fig.~\ref{fig:an}).

To verify approximations used in the analytical considerations, we compere them with the FEM results for the shallow channel of length $l\ll{}a$. 
Fig.~\ref{fig:fem_vs_an}(left) shows comparison between analytical (solid lines) and FEM results (dashed) for the $j_y$ component of the flow, taken along a cut through the channel (orange dashed line in Fig~\ref{fig:FEM}(top)). Small discrepancies can be observed due to the finite channel length, i.e., $l=a/20$. This difference can be controlled by adjusting $l$, although $l$ must remain nonzero in the FEM formulation. The analytical profiles are normalized so that their central values match the numerical ones. Moreover, our results for $\lambda=a/10$ match the experimental electronic-flow profile in graphene~\cite{jenkins2022imaging}, shown as the black dotted line in Fig.~\ref{fig:fem_vs_an}(left). To fit the experimental profile, we put $j_0=13$~A/m.

To verify the logarithmic behavior of the potential at a distance from the constriction, we also present -- in Fig.~\ref{fig:fem_vs_an}(right), FEM results for the potential $\varphi(L_x/2,y)$ along the section depicted in Fig.~\ref{fig:FEM}(top) as black dashed line. Logarithmic fits $-A\log\left(\frac{y}{\lambda}\right)-B$ (solid lines) are added to each $\varphi$ profile (dashed lines) showing approximate logarithmic behavior of the potential in a certain region of the computational box.    

\section{Finite channel length}
\begin{figure}[bt]
    \centering
    \includegraphics[width=0.99\linewidth]{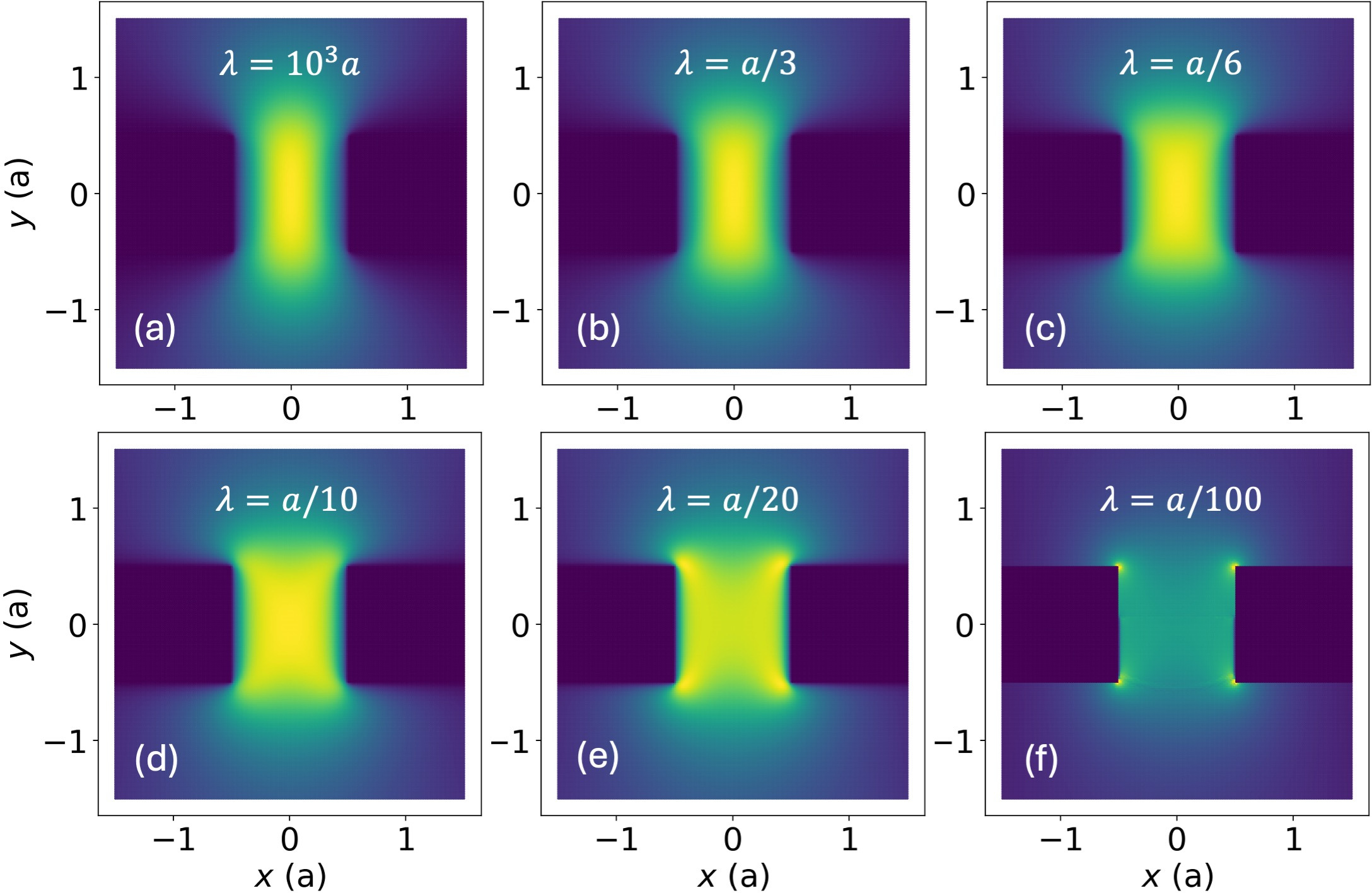}\\
    \includegraphics[width=0.89\linewidth]{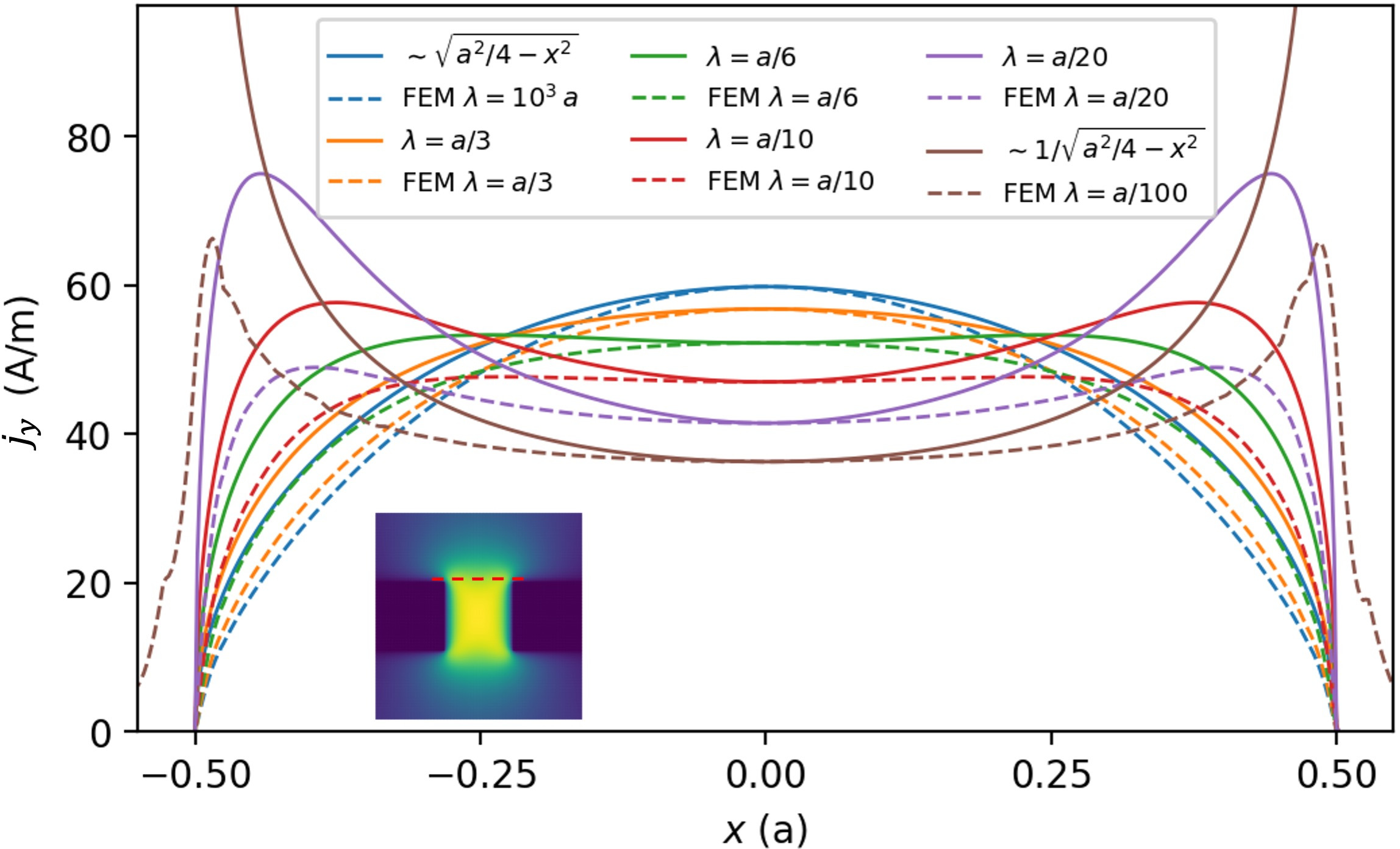}\\
    \includegraphics[width=0.89\linewidth]{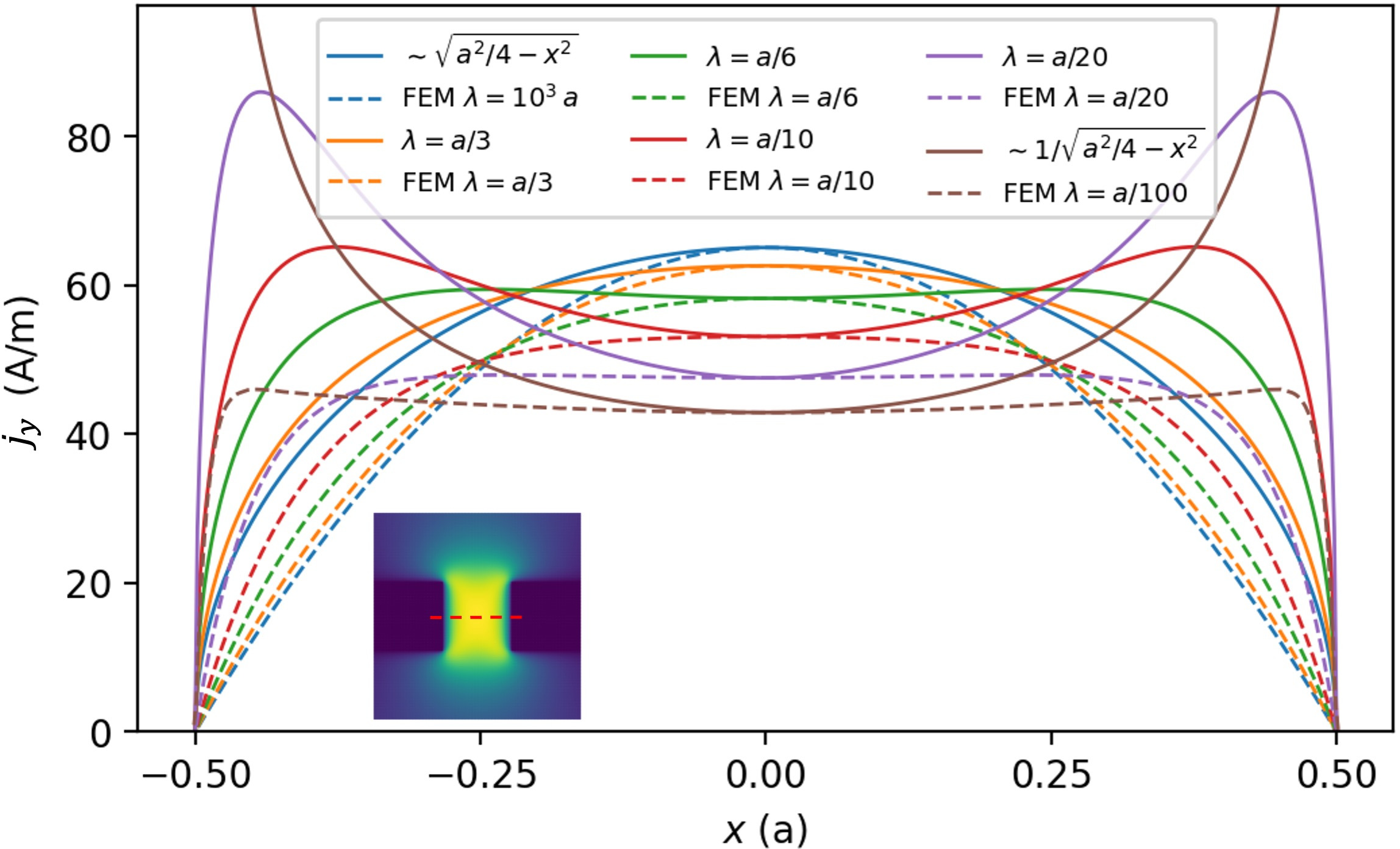}
    \caption{(top) FEM calculations showing the current density maps $|j|$ for longer channel (of length $l=a$) with increasing momentum relaxation: $\lambda=10^3a$, $a/3$, $a/6$, $a/10$, $a/20$, and $a/100$ for (a-f) subplots, respectively.
    Current $j_y$ component profiles across the channel aperture along two different cut sections: in the channel outflow (middle panel) and the channel middle (bottom panel), shown in respective insets as red dashed sections.}
    \label{fig:fem_vs_an_fins}
\end{figure}

We also introduced FEM to complete the analytical results and extend them on more complex geometries than infinite thin constriction, discussed in Fig.~\ref{fig:fem_vs_an}.
In case of finite channel lengths, for example $l=a$, the profiles are different and presented in Fig.~\ref{fig:fem_vs_an_fins}.
The first important difference is that the double-bump structure can be observed only in the channel outflow area -- see Fig.~\ref{fig:fem_vs_an_fins}(top). In the outflow area -- see Fig.~\ref{fig:fem_vs_an_fins}(middle) -- we observe transition from semicircular (for large $\lambda$) to double-bump (smaller $\lambda$) current density profiles. While in the middle of the channel aperture -- Fig.~\ref{fig:fem_vs_an_fins}(bottom) a Poiseuille (for large $\lambda$) or flat (smaller $\lambda$) current density profiles are observed. 
The FEM results (dashed lines) are no longer well reproduced by the analytical results (solid lines), especially in the middle part of the channel. At the opening, the numerical profiles exhibit a certain qualitative agreement with the analytical profiles, derived for infinitely thin constriction.

\section{Convective term}

\begin{figure*}[htb]
    \centering
\includegraphics[width=.45\linewidth]{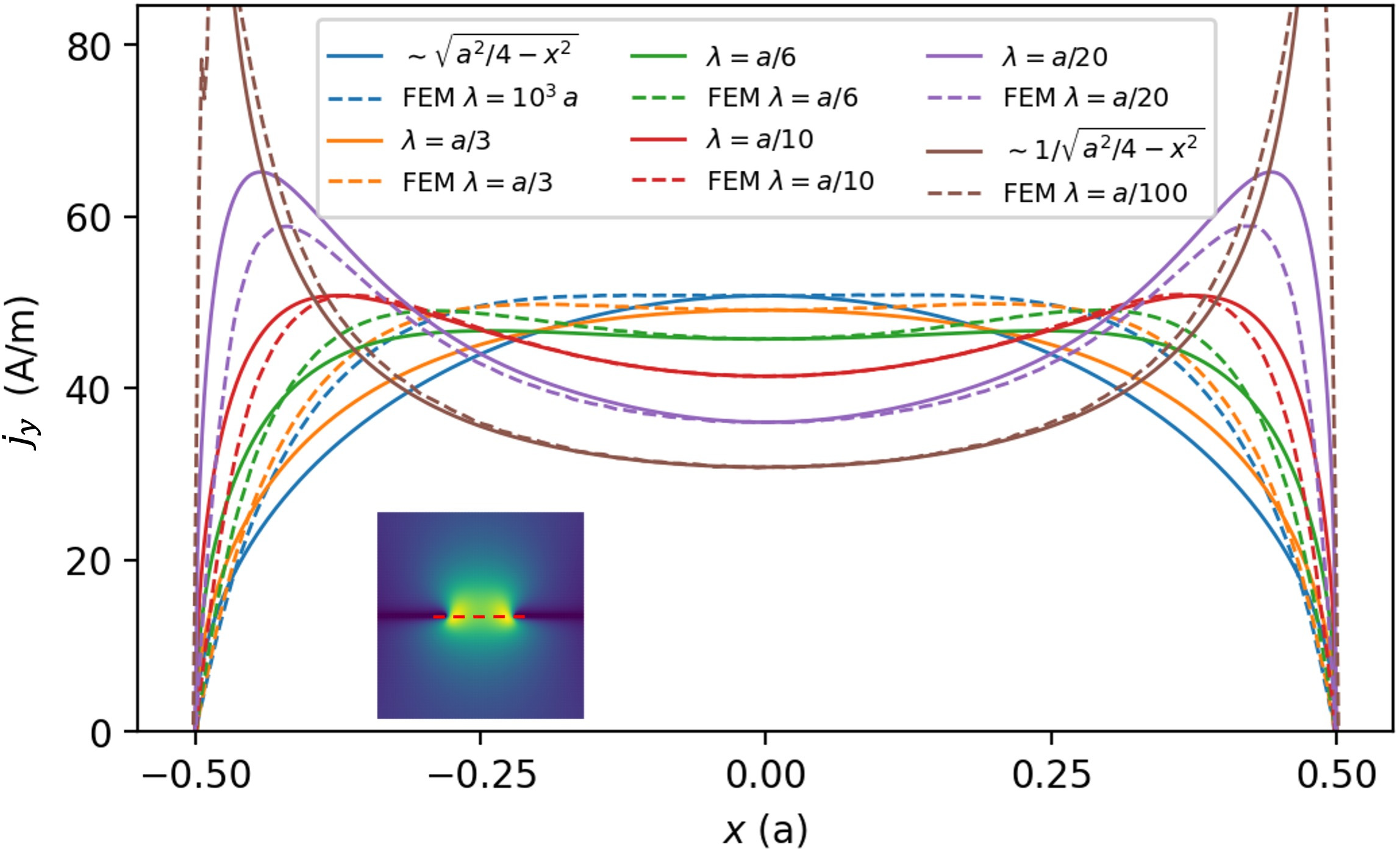}
\hspace{3mm}
\includegraphics[width=.37\linewidth]{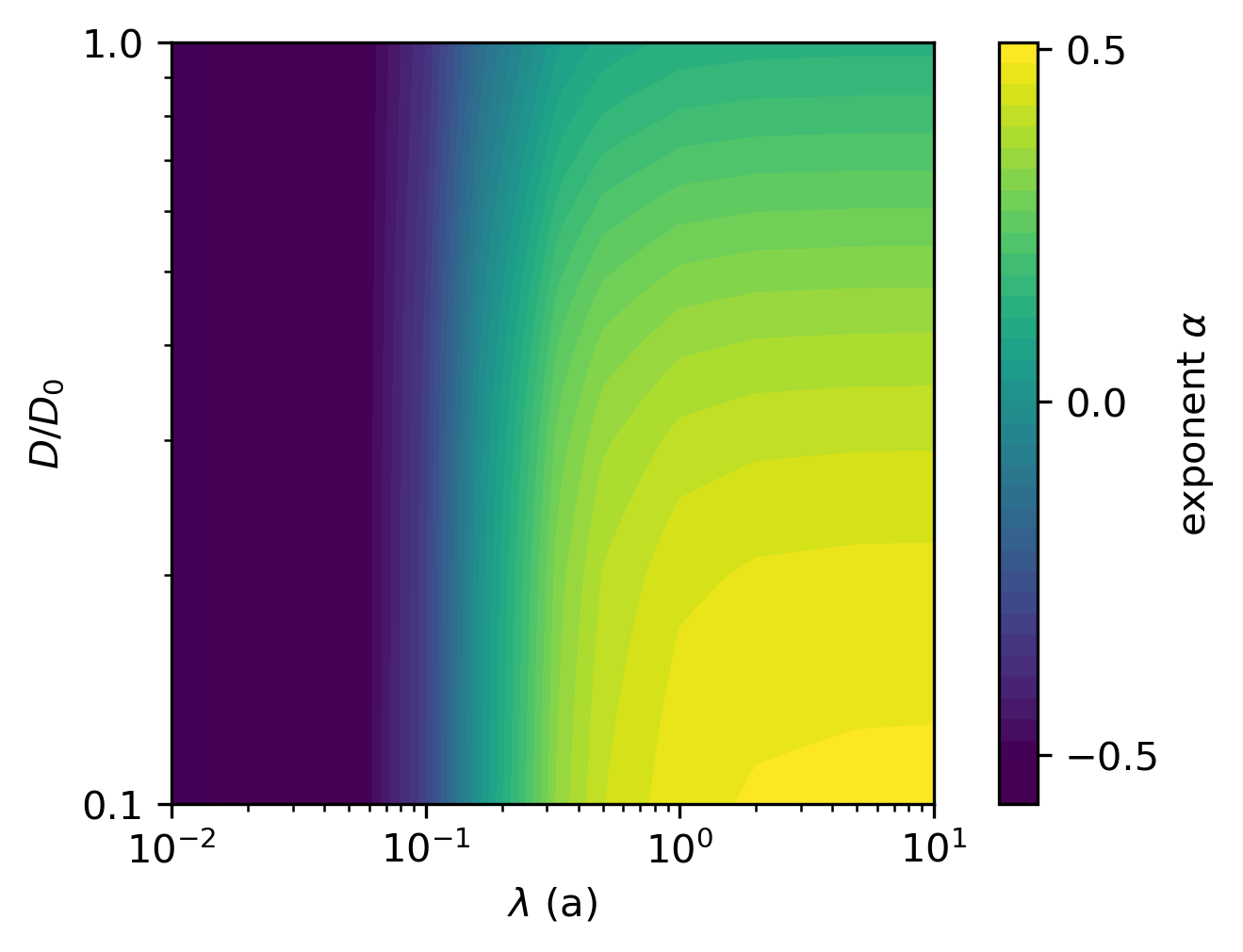}
    \caption{(left) Same as Fig.~\ref{fig:fem_vs_an} but with the convective term included in the FEM calculations.
    (right) Fitted $\alpha$ exponent, convective case.}
    \label{fig:fem_conv_s}
\end{figure*}

In Fig.~\ref{fig:fem_conv_s} we present similar results as in Fig.~\ref{fig:fem_vs_an} but now in the FEM formulation a convective term $\sigma D j_k\partial_kj_i$ in the RHS of Eq.~(\ref{eq:brinkman}) is included, where $D=\frac{\mu}{v^2_Fe^3n^2}$~\cite{Lucas2018}. For typical parameter values~\cite{jenkins2022imaging}: $n = 10^{12}\,\mathrm{cm^{-2}}$,
$v_F = 10^6\,\mathrm{m/s}$, $\mu \approx 0.12\,\mathrm{eV}$,
one finds $D\approx 4\times 10^{-8}\ \mathrm{m^2V/A^2}$.
For estimates $|\mathbf j| \sim 50\,\mathrm{A/m}$ and $|\nabla \mathbf j| \sim 10^7\,\mathrm{A/m^2}$, as in the graphene experiment presented in~\cite{jenkins2022imaging},
the convective electric field is
$E_{\mathrm{conv}} = D\, |\mathbf j||\nabla \mathbf j|\sim 20\ \mathrm{V/m}$,
while the Ohmic field is $E_{\mathrm{Ohm}} = |\mathbf j|/\sigma\sim 500\,\mathrm{V/m}$ (for $\sigma \simeq 0.1\,\mathrm{S/sq}$).  
Thus, the convective correction in graphene is of the order of a few percent of the Ohmic term.
In the simulations, however, to make convective effects visible, we assume the convective term to have the value $D=D_0=10^{-6}\ \mathrm{m^2 V/A^2}$, such that for $\lambda = a/10$ in the channel region $E_{\mathrm{conv}} \sim E_{\mathrm{Ohm}}$.

The main effect observed after adding the nonlinear convective term is the emergence of asymmetry in the current-density profile along the channel $j_y(y)$, as shown in Fig.~\ref{fig:fem_conv} in the main text. Moreover, if we examine the profiles $j_y(x)$ across the channel aperture in Fig.~\ref{fig:fem_conv_s}(left), the inclusion of nonlinear effects causes the flattened, non–Poiseuille-like structure to appear earlier---at larger values of $\lambda$ than in the case without convection. A similar behavior is observed in Fig.~\ref{fig:fem_conv_s}(right), where the exponent of the fitted $A\!\left(a^2-4x^2\right)^\alpha$ is shown.

\bibliography{main}
\end{document}